\documentclass[11pt, a4paper]{article}
\pdfoutput=1
\usepackage[utf8]{inputenc}
\usepackage[T1]{fontenc}
\usepackage{amsmath, amssymb, amsthm, amsfonts} % For Rank, Sigma, etc.
\usepackage{graphicx} % To include Figure 2
\usepackage{booktabs} % For Table 1 (Tableextraperf)
\usepackage{natbib}   % For \cite{laloux1999random}
\usepackage[hidelinks]{hyperref} % Clickable links without ugly boxes
\usepackage{geometry}
\usepackage{graphicx}
\usepackage{rotating}
\usepackage{adjustbox}
\usepackage{booktabs}
\usepackage{multirow}
\usepackage{epstopdf}

\theoremstyle{thmstyleone}%
\theoremstyle{thmstyletwo}%
\newtheorem{example}{Example}%
\newtheorem{remark}{Remark}%

\pdfoutput=1

\title{Measuring the risk or reducing it, that is the question: is risk measurement necessary for risk reduction?}
\author{Pierpaolo Uberti\\ \small University of Milano-Bicocca \\ \small \texttt{pierpaolo.uberti@unimib.it}}
\date{\today}

\begin{document}

\maketitle

\begin{abstract}
In this research, starting from a widely accepted definition of risk, we support the idea that risk reduction is a more realistic objective than risk minimization, which represents a theoretical utopia. Furthermore, significant risk reduction can be achieved without relying on risk measurement and risk minimization. To this end, we propose a generalization of the numerical rank and the condition number of a matrix, specifically the return matrix in this application. This generalization considers the entire matrix spectrum instead of focusing only on the smallest eigenvalue, as the condition number does. The approach directly provides an order among a finite number of risky scenarios. Risk reduction is obtained by identifying the riskiest scenarios and reducing investment exposures corresponding to them. The validity of this theoretical proposal is supported by a comprehensive experiment performed on real data. The capacity of the proposed approach to effectively reduce risk is proven by measuring the variability of out-of-sample returns for benchmark portfolios-constructed by minimizing standard risk measures-compared to the strategy of reducing exposure in high-risk scenarios. Finally, preventing large losses with limited active management-thereby controlling the impact of transaction costs-not only reduces risk but also preserves the average return and, consequently, the portfolio's Sharpe ratio.
\end{abstract}


\section{Introduction}\label{intro}

Markowitz's work (see \cite{M}) laid the foundations of modern portfolio theory. For mathematical convenience, the original formulation of the allocation model measures the risk of a portfolio by the variance of its return. Markowitz's general contribution can be summarized and generalized as follows: given a specific risk measure, the optimal investment portfolio is the one that minimizes risk subject to two primary constraints: the budget constraint (ensuring the solution is a feasible portfolio) and a constraint on the portfolio's expected return. 

In both practical applications and theoretical research, numerous risk measures have been proposed. A first systematic approach to risk measures dates back to the seminal work of Artzner et al. (see \cite{artzner1999coherent}), where the authors introduced an axiomatic definition of coherent risk measures. A risk measure is coherent if it satisfies four minimal properties: monotonicity, translation invariance, subadditivity, and positive homogeneity. Following this paper, an important stream of literature emerged proposing alternative axiomatic definitions: among others, we recall \cite{follmer2002convex} and \cite{frittelli2002putting} for convex risk measures, \cite{acerbi2002spectral} for spectral risk measures, \cite{rockafellar2006generalized} for deviation risk measures, and \cite{acciaio2011dynamic} for dynamic risk measures.

The paper by \cite{artzner1999coherent} also provides a broad and widely accepted definition of risk: ``risk is related to the variability of the future value of a position, due to market changes or more generally to uncertain events''. We define risk according to this perspective. The crucial point of the definition is that risk is concerned with the variability of \textsl{future} returns. Concisely, risk mitigation is about reducing, or at least managing, uncertainty. This implies that a risk measure, when computed based on past returns, essentially reduces to a measure of data variability. To distinguish a risk measure from a mere variability measure, it is necessary to predict future returns, unless one implicitly assumes that past returns are a good predictor of future returns. This is one of the main reasons why risk measures often fail to prevent large future losses in practice, see \cite{danielsson2008blame} and \cite{danielsson2016model}. Risk management is, ultimately, all about forecasting.

Although often not explicitly or sufficiently emphasized, risk measures assume measurability in probabilistic terms. This means that, when using a risk measure, it is assumed that both the set of possible events and their probabilities are known. As highlighted in several papers (see, e.g., \cite{mousavi2014risk}), this hypothesis is valid in controlled laboratory experiments, such as rolling a die, where the set of outcomes and their probabilities are precisely defined, or in lotteries. By contrast, in real-world applications, and particularly in financial investments, it is often challenging to even identify the relevant set of events, let alone assign probabilities to them.

An additional issue arises from the fact that the most severe and catastrophic events are typically associated with extremely low, or even negligible, probabilities, see \cite{kelly2014tail}, \cite{barro2021rare}, \cite{daniels2026economic} and \cite{oecd2026catastrophic}. The rarity of such events does not preclude their occurrence; hence, relying solely on probabilistic measures may lead to an underestimation of exposure to extreme risks. The key issue is that measuring risk entails a ``great expectation'', an endeavor far broader and more complex than the primary objective of controlling or reducing future variability. This does not imply that attempting to measure risk is meaningless. Different risk measures emphasize different aspects of risk: for instance, Value-at-Risk (VaR) and Expected Shortfall (CVaR) focus asymmetrically on the left tail of the return distribution, whereas others, such as standard deviation or mean absolute deviation, capture overall variability symmetrically. Risk measures are also essential in regulatory contexts; for example, the computation of capital requirements for bank investments prescribes the use of CVaR under Basel III guidelines. Moreover, risk measures serve as rigorous mathematical instruments for comparing the performance of alternative investment strategies.  

The key result supported in this paper can be summarized as follows: both academic researchers and practitioners should accept the evidence that the most effective way to achieve risk reduction is not necessarily to choose the allocation that minimizes a single given risk measure. While such an approach may sound meaningful and natural in theoretical settings, it does not consistently lead to good results in practice. A dedicated section of this paper provides a comprehensive discussion of the shortcomings inherent in classical allocation models, alongside the various solutions proposed in the existing literature; for a detailed analysis, the reader is referred to Section \ref{lit}. To clarify the logic underlying the novel proposed perspective, it is useful to consider a real-life example where risk is both concrete and potentially severe.

Suppose one is walking along the Magnificent Mile during a cold winter in Chicago, aware of the possibility, however remote, that a piece of ice might fall from a skyscraper and cause fatal injury. In such a situation, the rational response is simply to walk away from the buildings, without attempting to calculate the probability of such an event. That probability is extremely small, comparable to being struck by lightning, yet the severity of the potential outcome renders the situation unacceptably risky. Now, one might argue that by walking farther from the buildings, a pedestrian moves closer to the street, thereby increasing the probability of being hit by a taxi skidding on the ice. This trade-off, though somewhat anecdotal, captures the essence of the issue: in real life, it is already challenging to identify all sources of risk, let alone assign probabilities to them. It is often sufficient, and indeed more effective, to recognize hazardous scenarios and avoid them altogether, irrespective of their perceived likelihood.

This reasoning underpins the methodological approach proposed in the present paper. Despite the illustrative analogy used to introduce the concept, this research provides a formalized and rigorous mathematical framework. The identification of high-risk scenarios is the result of a suitable generalization of matrix conditioning techniques. If diversification is the secret behind risk reduction, a point that will be supported and argued throughout this paper, then the rank of the return matrix for a given investment universe should serve as a highly informative indicator of the number of effective diversification opportunities. The idea of considering the number of effective available diversification opportunities is a concept already studied in the literature, see among others \cite{meucci2009managing}, \cite{meucci2013measuring} and \cite{bera2008optimal}. 

One might assume that the rank of a matrix and related concepts are of limited utility since they only capture linear dependencies between assets. Nevertheless, linear dependence is exhaustive in a framework where an investor can only combine assets into portfolios, which are themselves linear combinations of assets. To measure risk and identify high-risk scenarios, we propose a generalization of the condition number and the numerical rank of a matrix. A generalization of these standard concepts is necessary because the exact rank, calculated in an algebraic sense, carries negligible informative content in this context, as will be argued. From a numerical stability perspective, the smallest singular value of a matrix measures the distance to the closest rank-deficient matrix, see \cite{golub2013matrix}. By construction, if the reference matrix has rank $N$, the rank of the closest rank-deficient matrix is $N-1$. Since most standard quantitative applications require full-rank invertible matrices, it is usually sufficient to consider only the distance to the closest rank-deficient matrix. However, in risk applications, it is more insightful to use the numerical rank and consider the distances to rank-deficient matrices of rank $N-2$, $N-3$, and so on. To this end, the entire spectrum of the matrix must be considered. Depending on the distances to matrices of different ranks, a finite set of possible risky scenarios, ordered from the least to the most risky, can be defined. Finally, a straightforward trading strategy is implemented: invest in a given portfolio when the risk is manageable, and reduce exposure when high-risk scenarios are identified. This strategy is deliberately parsimonious to better test the forecasting power of the proposed approach. Furthermore, it remains agnostic regarding portfolio composition, as our framework indirectly supports the evidence that specific allocation is of secondary importance for risk reduction compared to the identification of high-risk scenarios. Finally, the effectiveness of the proposed approach in reducing risk (and often preserving the portfolio's Sharpe ratio) is tested through comprehensive experiments performed on several datasets of real financial data. We compare the out-of-sample returns of our approach with those of standard allocations based on the minimization of a single risk measure.

The paper is organized as follows. Section \ref{lit} summarizes the general discussion framework regarding the shortcomings of classical allocation models. Section \ref{idea} presents the theoretical framework along with its underlying assumptions. In Section \ref{theory}, the methodology is formalized and illustrated with graphical examples. Section \ref{app} details the empirical application of the trading strategy based on the identified risky scenarios using a primary database. Finally, Section \ref{concl} concludes the paper, while the Appendix \ref{app} reports the results for the additional databases.

\section{The debate on the criticism of classical asset allocation models} \label{lit}

The present paper proposes a novel and critical approach to risk reduction as an alternative to the standard methodologies found in the literature. The critique is not merely directed at the measurement of risk itself, but also at the reliance on optimization frameworks to determine asset allocations that minimize a specific risk measure. Consequently, this study contributes to the extensive debate surrounding asset allocation models and their out-of-sample performance. This section does not intend to provide an exhaustive literature review, which is beyond the scope of this research, but rather offers a concise overview of the central themes in this academic debate.

While the work of Markowitz \cite{M} remains the foundational reference for Modern Portfolio Theory, a vast body of literature has highlighted the significant practical challenges and poor out-of-sample results associated with mean-variance optimization. A primary critique is provided by \cite{michaud1989markowitz}, who explains that traditional risk minimizers treat point estimates of risk and return as absolute truths, entirely ignoring statistical uncertainty. This results in counter-intuitive and highly unstable portfolio weights, a phenomenon often described as ``error maximization''.

In a seminal empirical study, \cite{DeMig} demonstrated that none of the sophisticated optimization models consistently outperform a naive, equally weighted (1/N) portfolio out-of-sample in terms of Sharpe ratio, certainty-equivalent return, or turnover. They conclude that the theoretical gains from optimal diversification are typically negated by parameter estimation errors and transaction costs. Similarly, \cite{jagannathan2003risk} show that minimum-variance portfolios often perform poorly because the sample covariance matrix is difficult to estimate accurately for large investment universes. Interestingly, they find that imposing theoretically ``arbitrary'' constraints, such as forbidding short sales, actually improves performance by acting as a shrinkage penalty that limits the impact of estimation error.

Furthermore, \cite{chopra1993effect} provide evidence that errors in estimating expected returns are roughly ten times more damaging than errors in variances and twenty times more damaging than errors in covariances. This explains why minimum-variance portfolios, which ignore expected returns, often outperform full mean-variance models. It should be noted that the approach presented in this paper is strictly risk-oriented; therefore, expected returns are excluded from all compared allocation models. Finally, \cite{kan2007optimal} document that simply ``plugging in'' sample estimates leads to massive performance degradation, suggesting that parameter uncertainty is the primary obstacle to successful risk-based allocation.

Following these milestones, the academic debate has evolved in several directions which, upon deeper analysis, represent different perspectives on the same fundamental problem. A substantial body of research \cite{gelmini2024equally, malladi2017equal, zakamulin2017superiority, yuan2024naive} continues to confirm the superiority of naive allocation over complex optimization in out-of-sample tests. Conversely, some authors defend the optimization framework \cite{demiguel2009generalized, allen2019defense, kritzman2010defense}, often emphasizing that the success of these models is contingent upon the accuracy of the underlying forecasts.

One major stream of research focuses on Shrinkage Estimators, which mitigate estimation error by regularizing the sample covariance matrix toward a more stable target structure \cite{ledoit2004well, ledoit2003improved}. These methodologies have evolved from early linear models to sophisticated nonlinear approaches \cite{ledoit2017nonlinear, ledoit2022power}. Others address Robust Optimization, which explicitly incorporates parameter uncertainty by assuming that true parameters lie within a predefined ``uncertainty set'' \cite{kim, xidonas2020robust, ghahtarani2022robust}. By seeking a portfolio that performs best under a ``worst-case'' scenario, these models attempt to shield the allocation from the extreme sensitivity typical of traditional optimizers.

The 2008 global financial crisis further exposed the inability of standard models to handle extreme market conditions, leading to the rise of new paradigms, risk parity and maximum diversification. Risk Parity shifts the focus from capital allocation to risk allocation, ensuring each asset contributes equally to the total portfolio risk \cite{maillard2010properties, roncalli2013introduction}. This framework has been extended to various risk measures, including VaR \cite{bruder2012managing}, Expected Shortfall \cite{mausser2018long}, and Mean Absolute Deviation \cite{ararat2024mad}. Diversification maximization, rather than minimizing a risk proxy, aims to maximize the portfolio's ``breadth''. Strategies include optimizing the Diversification Ratio \cite{choueifaty2008toward}, maximizing the effective number of independent risk factors \cite{meucci2009managing}, or utilizing entropy-based measures \cite{yu2014diversified}. The literature on diversification measurement in portfolio is wide and variegated; for a review see \cite{lhabitant2017portfolio} and \cite{koumou2020diversification}.

Despite these advancements, most existing models still focus on the internal distribution of wealth. The methodology proposed in this paper departs from this tradition by suggesting that the detection of structural changes in the return matrix, and the subsequent timing of leverage, is more critical for risk reduction than the specific allocation weights themselves.

\section{The idea behind the approach}\label{idea}

Let us consider a market with $N$ risky assets. The return time series of these assets form the columns of a $T \times N$ matrix $A$, where $T$ represents the length of the time series. The rank of $A$ satisfies $\operatorname{rank}(A) \leq \min \{T,N\}$. In standard financial applications, it is typical to assume $T \geq N$. When working with empirical data, it is statistically improbable for the return of one asset to be an exact linear combination of the returns of the other assets. Consequently, the assumption $T \geq N$ almost always implies that $\operatorname{rank}(A) = N$. From a risk perspective, the algebraic rank of $A$ is therefore an ineffective risk indicator, as it remains constant at $N$ whenever $T \geq N$.\footnote{The rank calculated in an analytical sense is equally uninformative when $T < N$ because, for the same reasons, $\operatorname{rank}(A) = T$.} Accordingly, and without loss of generality, we assume for the remainder of this paper that the return matrix $A$ has full column rank $N$.

The objective is to define a suitable generalization of the numerical rank of a matrix, specifically the return matrix A, to serve as a robust risk indicator. During financial crises, losses are typically severe and widespread across the majority of market assets. This phenomenon implies that asset correlations surge, rendering risk reduction through diversification less effective, if not entirely impossible. This phenomenon, often termed a diversification meltdown, has been empirically documented in recent market turmoil where traditional asset hedges failed to provide protection (see, e.g., \cite{imf2026diversification, ecb2025connectedness}). Since the algebraic rank of the return matrix is almost invariably $N$, one must transition from analytical to numerical linear algebra to extract meaningful information regarding the diversification opportunities at risk of being lost. 

One immediate possibility is to compute the numerical rank rather than the exact rank. The numerical rank serves as a potential indicator of the effective dimensionality of the investment universe; a decrease in this rank signals a reduction in available diversification opportunities. Previous applications of these concepts in the field of systemic risk can be found in \cite{maggi2020proper}, \cite{bartesaghi2025global}, \cite{pastorino2024empirical}, and \cite{figini2020market}. 

The present research takes a further step in generalizing numerical rank and matrix conditioning as risk indicators. In particular, this paper addresses a primary limitation of the numerical rank: the necessity of choosing an arbitrary threshold to determine which singular values are significantly different from zero. This choice is not merely a technicality. In standard numerical applications, the threshold is often set to machine precision to prevent computational instability, see \cite{golub2013matrix}, \cite{trefethen1997numerical} and \cite{higham2002accuracy}. However, in risk applications, such a threshold is often too conservative, yielding information as uninformative as the algebraic rank. Our approach is based on the intuition that critical risk information is hidden within the ``gray zone'' between algebraic and numerical rank, and is specifically designed to investigate this space.

The singular values of $A$ are denoted by $\sigma_1 < \sigma_2 < \sigma_3 < \dots < \sigma_N$, representing the \textit{spectrum} of $A$.\footnote{Strictly speaking, the spectrum of a matrix is defined as the set of its eigenvalues. Given the direct relationship between singular values and eigenvalues, we use this term here as a slight abuse of notation to identify the set of singular values.} For convenience and without loss of generality, we refer to the normalized spectrum of a matrix, $\overline{S}(A)$, as the vector containing the singular values divided by the largest singular value, $\sigma_N$. Thus, the normalized spectrum is defined as:
$ \overline{S}(A) = [ \overline{\sigma}_1 = \frac{\sigma_1}{\sigma_N}, \quad \overline{\sigma}_2 = \frac{\sigma_2}{\sigma_N}, \quad \dots, \quad \overline{\sigma}_{N-1} = \frac{\sigma_{N-1}}{\sigma_N}, \quad \overline{\sigma}_N = 1 ]$. In this framework, each return matrix $A$ is identified by a vector of dimension $N-1$, whose entries are the normalized singular values. Since the maximum singular value of the normalized spectrum is equal to 1 for all matrices, it is omitted as it provides no discriminatory information. 

It is straightforward to observe that while a matrix is described by its normalized spectrum, the converse is not true; given a set of $N-1$ real numbers in the interval $[0,1]$, infinitely many matrices may share the same normalized spectrum. The minimum value of the normalized spectrum is directly related to the reciprocal of the condition number of the reference matrix. As is well known (see \cite{golub2013matrix}), this value represents the distance, with respect to the Euclidean norm\footnote{The choice of a specific distance metric does not affect the underlying logic of the proposed approach. Unless otherwise specified, we utilize the Euclidean distance.} to the nearest rank-deficient matrix. By construction, the closest rank-deficient matrix to a matrix of rank $N$ is always a matrix of rank $N-1$. Consequently, the condition number focuses exclusively on the loss of a single dimension of rank. The wide use of the condition number in standard literature is intuitive: when full-rank matrices are required to ensure the existence of an inverse (or a pseudo-inverse, as in Ordinary Least Squares estimation), the loss of even one dimension is sufficient to preclude further computation. However, a further potential issue arises: the closest rank-deficient matrix does not necessarily inherit the structural properties of the original matrix. Therefore, the comparison between the original matrix and its closest rank-deficient counterpart may be inconsistent. For instance, while a covariance matrix is symmetric and positive definite, its closest rank-deficient matrix is not guaranteed to satisfy either property. In such cases, the closest rank-deficient matrix cannot represent a valid covariance matrix, further limiting the informative value of the usual approach.

The rationale behind the proposed approach is as follows. The fact that $\operatorname{rank}(A) = N$ and that the (Euclidean) distance to the nearest matrix of rank $N-1$ is measured by $\overline{\sigma}_1$ provides only limited insight into the risk of investing in a portfolio of the $N$ given assets. A more robust perspective for risk detection would consider the distances to the nearest rank-deficient matrices of rank $N-2, N-3, \dots$. Such an approach generalizes the concept of numerical rank, capturing potentially critical risk information. If a return matrix $A$ is close to a rank-deficient matrix of rank $N-1$ (which is always the closest rank-deficient matrix by construction) but remains relatively distant from matrices of rank $N-2$, $N-3$, and so on, the scenario is not particularly alarming. Conversely, a matrix $A$ might be very close not only to a matrix of rank $N-1$, but also to rank-deficient matrices of rank $N-2, N-3, \dots$. To use a geometric analogy: the first case is akin to standing on the edge of a single step, where the risk is limited to falling one level lower. The second case is like standing on the edge of a high cliff, where a potential fall would be far deeper and more catastrophic. Translating this intuition into a financial risk context, we support the idea that it is fundamental to distinguish between a potential loss of a single dimension and a simultaneous collapse of multiple dimensions. In the latter case, numerous diversification opportunities vanish at once, rendering the $N$-asset universe highly risky due to the lack of effective opportunities for risk reduction.

\begin{remark}\label{rem1}
Coherent measures of risk, as defined in the seminal work of \cite{artzner1999coherent}, are characterized by four fundamental properties. \textsl{Non-negativity} is a technical requirement ensuring that risk values remain non-negative, facilitating interpretation. \textsl{Translation invariance} describes the impact of a deterministic shift in returns, such as the inclusion of a risk-free asset in the investment universe. These two properties do not provide a functional basis for risk-reduction strategies.
\textsl{Positive homogeneity} governs scale transformations, dictating how portfolio leverage affects risk; specifically, while the absolute magnitude of risk scales linearly with leverage, the risk-per-unit-of-leverage remains constant. This property highlights a trivial, yet often underestimated, mechanism for risk management: since absolute gains and losses depend entirely on the level of exposure, risk can be adjusted simply by scaling leverage, a process requiring no forecasting ability. Critically, only \textsl{subadditivity} provides a structural mechanism for risk mitigation. Formulated as an inequality, this property states that the risk of a portfolio is never greater than the sum of the risks of its individual components. Therefore, for any subadditive risk measure, it is a tautology that diversification reduces risk. From this, we conclude that if diversification, formalized through subadditivity, is the primary driver of risk reduction, then the fact that a return matrix typically possesses full rank $N$ in a strictly algebraic sense is a misleading risk indicator. Such a static, binary view of rank inherently leads to an underestimation of risk. What truly determines the stability and risk-reduction potential of an allocation is the distance to the nearest rank-deficient matrices and the potential magnitude of that rank loss.
\end{remark}

In the following section, we formalize a methodology to derive a set of ordered risky scenarios based on the normalized spectrum of the return matrix. These scenarios represent varying degrees of potential dimensionality loss, providing a structured framework to quantify the risk previously discussed. The investment strategy proposed in the appication effectively synthesizes the risk-reduction prerogatives of both positive homogeneity and subadditivity. Our geometric intuition regarding the potential loss of effective diversification opportunities is utilized to forecast heightened risk levels. Subsequently, tangible risk reduction is achieved by dynamically adjusting the portfolio's leverage, and thus its total exposure, in direct correspondence with these identified scenarios. By reducing exposure precisely when the ``effective'' dimensionality of the investment universe collapses, the strategy mitigates potential catastrophic losses that standard optimization models, often fail to anticipate.

\section{The theoretical proposal}\label{theory}

Let us define $\mathcal{M}_N$ as the set of matrices of rank at most $N$. Thanks to the assumptions in Section \ref{idea}, each return matrix $A$ is an element of $\mathcal{M}_N$ and can be represented as a point in $\mathbb{R}^{N-1}$; the coordinates of this point are the $N-1$ entries of $\overline{S}(A)$.\footnote{As assumed in Section \ref{idea}, without loss of generality, the elements of the normalized spectrum are collected in strictly ascending order. Let us underline that the order of the elements of the normalized spectrum does not impact the application of the proposal. Moreover, while it is theoretically possible for two or more singular values to be equal, in practical applications this occurrence has null probability and the assumption is always verified.} Moreover, since the elements of $\overline{S}(A)$ belong to the real interval $(0,1)$, each matrix in $\mathcal{M}_N$ is identified by a point belonging to the unit hypercube in $\mathbb{R}^{N-1}$ with vertices given by all vectors in $\{0,1\}^{N-1}$. In such a representation, the matrices of full rank $N$ and condition number strictly larger than $1$ cover all the internal points of the hypercube. The matrices with rank $N$ and unitary condition number, all the proportional transformations of the identity matrix, correspond to the vertex $[1,\dots,1]$. The rank-deficient matrices correspond to points belonging either to the vertices (except for the vertex $[1,\dots,1]$), faces, or edges of the unit hypercube. The origin of the axes corresponds to a return matrix with unit rank; it represents the worst scenario in terms of rank and, indeed, in terms of diversification opportunities that are lost. The above representation is coherent with the standard result for which the closest rank-deficient matrix to a matrix of full rank $N$ is of rank $N-1$ and the distance is measured by the smallest singular value. Indeed, given an interior point of the hypercube (a matrix of rank $N$), if starting from the ordered normalized spectrum, the closest rank-deficient matrix always belongs to the face with coordinates $[0,\cdot,\dots,\cdot]$. A graphical representation to support the geometric intuition is reported in Example \ref{es1}.

Starting from a given set of $N$ assets and a corresponding return matrix $A$, we propose to place the matrix in the unit hypercube of $\mathbb{R}^{N-1}$ by plotting the point identified by the normalized spectrum $\overline{S}(A)$. The first interesting fact to notice is that infinitely many internal points of the unit hypercube are characterized by the same condition number, even if their positions in the hypercube are very different. The different positions in the space could unveil significant differences in terms of potential loss of rank. Nevertheless, they are indifferent in terms of condition number. This evidence further supports the idea that the condition number, depending exclusively on the smallest singular value, could hide some relevant information in this context. The following example graphically highlights the intuition.

\begin{example}\label{es1}
To ensure a graphical representation, a toy example with $N = 4$ risky assets is considered. The normalized spectra of three different return matrices $A, B$, and $C$ are collected in the following vectors:
\[
\overline{S}(A) = [0.02 \quad 0.85 \quad 0.9]
\]
\[
\overline{S}(B) = [0.02 \quad 0.021 \quad 0.9]
\]
\[
\overline{S}(C) = [0.02 \quad 0.021 \quad 0.0215].
\]
The three matrices are characterized, by construction, by the same condition number. When the points are drawn in $\mathbb{R}^3$ together with the unit cube, it becomes graphically evident that the three matrices correspond to very different positions in the space, see Figure \ref{fig1}.

\begin{figure}[h!]
\centering
\includegraphics[width=0.5\linewidth]{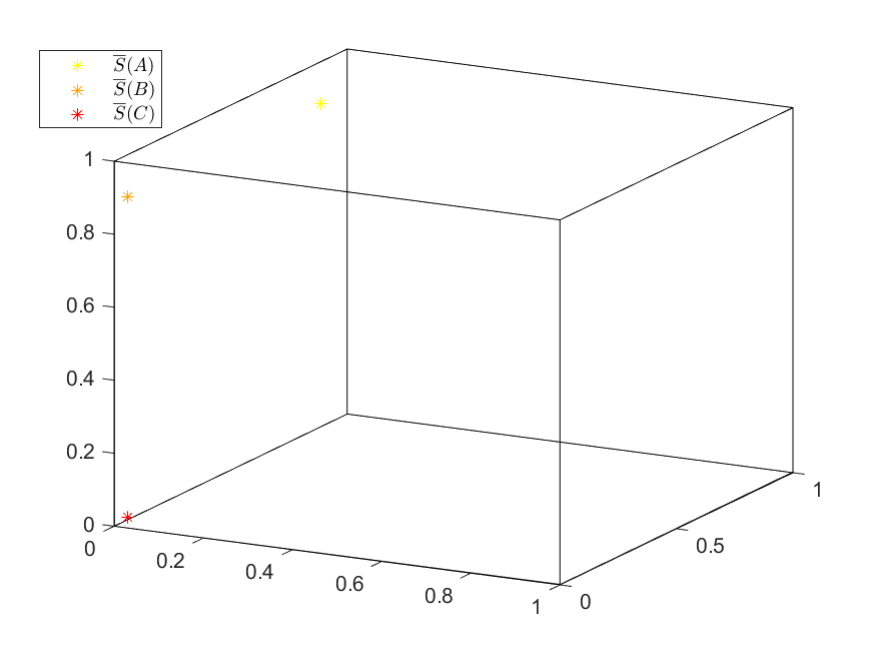}
\caption{The unit cube in $\mathbb{R}^3$ and the points representing matrices $A$, $B$ and $C$.}\label{fig1}
\end{figure}

The fact that the three matrices have the same condition number means that they are at the same (Euclidean) distance from the closest rank-deficient matrix, which is of rank $3$ in this example. Note that, since the elements of the normalized spectrum are collected in strictly ascending order, for each of the three matrices the closest rank-deficient matrix is represented by a point belonging to the left face of the unit cube. As already underlined, the order of the elements in the normalized spectrum does not impact the approach. What is fundamental to notice is that the three matrices are close to different vertices of the hypercube, each of them corresponding to a rank-deficient matrix of different rank. The number of ones in the coordinates of the vertices immediately returns the rank of the corresponding matrix.
  
\end{example}

The proposal consists in evaluating the different positions within the internal points of the hypercube to provide an interpretation in terms of risk. Since it is useless to calculate the rank of the closest rank-deficient matrix to discriminate different risky situations (see Example \ref{es1}), we propose to compute the distances to each of the vertices of the hypercube.\footnote{The choice of the distance does not qualitatively affect either the results nor the underlying idea.} The next step requires ordering the possible scenarios (represented by the distances to the vertices of the hypercube) from the least risky to the riskiest. One first elementary proposal is to discriminate between risky scenarios only considering the rank associated to the closest vertex. In such a framework, the riskiest scenario occurs when the closest vertex is the origin of the axes. The second riskiest scenario occurs when $[0, \dots, 0, 1]$ is the closest vertex. The third riskiest scenario occurs when $[0, \dots, 0, 1, 1]$ is the closest vertex. The least risky scenario occurs when $[1, \dots, 1]$ is the closest vertex. This classification rule is consistent with the geometric representation shown in Figure \ref{fig2}.

\begin{remark}
Without loss of generality, we have assumed that the elements of the normalized spectrum are in ascending order. Therefore, the number of meaningful vertices to be considered reduces from $2^{N-1}$ to $N$. In this framework, only the number of zeros and ones identifying a vertex are of interest.
\end{remark}

To further clarify the idea, a graphical example for $N = 4$ is provided. The internal points of the unit hypercube are drawn with different colors to support a direct intuition on the risk associated with different positions.
 
\begin{example}
Let us consider $N = 4$ risky assets. For simplicity, 100 points in $\mathbb{R}^3$ with coordinates in the interval $(0,1)$ are uniformly sampled. Then, following the idea expressed above, the different levels of risk related to the position of the points in the unit cube are represented with $4$ colors: green, yellow, orange, and red are used in this order to represent the risk scenarios, from the lowest risky to the highest one.
 
\begin{figure}[h!]
\centering
\includegraphics[width=0.5\linewidth]{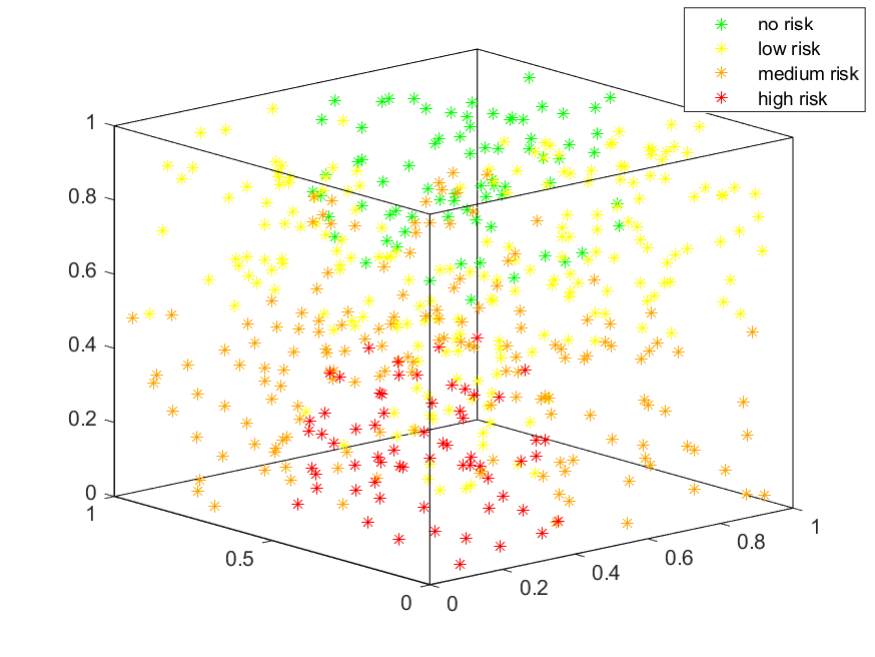}
\caption{Uniformly sampled points in the unit cube. According to the distance to the closest vertex, the different colors highlight the level of risk associated to each point.}\label{fig2}
\end{figure}

Figure \ref{fig2} provides immediate graphical information about the risk associated with different points in the hypercube. The closest vertex to the green points is $[1 \quad 1 \quad 1]$; for the yellow points, the closest vertex has one null coordinate. The vertex closest to the orange points is identified by two null coordinates, while the red points have the origin as the closest vertex. For the sake of graphical readability, the coordinates of the points are not ordered as assumed, so that the points uniformly cover the entire internal volume of the hypercube.

\end{example}

The conceptual framework established above and illustrated in the preceding examples is deliberately foundational and open to various refinements. For instance, one could consider not only the distance to the nearest vertex but also a simultaneous evaluation of distances to multiple vertices. The central challenge lies in choosing between a straightforward approach that maintains a naturally ordered set of risk scenarios and a more sophisticated one that may sacrifice this inherent ordering. In this context, the trade-off between simplicity and sophistication is a fundamental consideration. Within a forecasting framework, simpler models often prove more robust and effective than their complex counterparts, a critical observation given that risk management is fundamentally a forecasting exercise. This choice is grounded in the well-established ``Keep It Sophisticatedly Simple'' principle in forecasting \cite{zellner2001keep}, and is supported by extensive empirical evidence showing that simple models often outperform complex ones in out-of-sample settings \cite{makridakis1979accuracy, green2015simple}, particularly in the context of financial portfolio management \cite{DeMig}. Guided by this principle, the subsequent application is implemented with minimal sophistication. While more refined iterations of this basic idea may potentially yield superior results, such ad hoc refinements often fail to provide stable out-of-sample performance improvements, as they are frequently over-fitted to specific datasets or experiments.

\section{Application}\label{app}

In this section, we apply a rudimentary investment strategy based on the ideas expressed in the previous sections to test its performance. The strategy is uniquely designed to stress the risk-detection capabilities of the proposed approach.

We begin by describing the datasets and the general structure of the application. The numerical experiment is conducted on three distinct databases of daily returns spanning from January 5, 2005, to September 17, 2020. These databases are identified by their corresponding financial indices: S\&P500 $(n=378)$, Nikkei $(n=199)$, and FTSE $(n=79)$, where $n$ represents the number of stocks in each respective database. It should be noted that $n$ is smaller than the nominal number of index constituents because the application is restricted to assets with a complete time series over the entire observation period. We emphasize that the experiment is designed to be as general as possible; consequently, the results are independent of the specific asset sampling, the reference period, and the observation frequency.

The experiment is implemented through the following procedure. First, an investment universe of $N$ assets is randomly selected from one database. A standard rolling window exercise is then conducted on these $N$ assets using a fixed window length $w$. The alternative investment strategies, detailed below, are implemented by estimating parameters over the rolling window to compute the allocation used for the one-day-ahead out-of-sample returns. The window is then advanced by one day, and the process is repeated. To rigorously evaluate the robustness of the findings, the experiment is performed across multiple configurations of $N$ and $w$. Specifically, we test all combinations of N={5,10,20} and w={20,30,40}.\footnote{While these values for $N$ and $w$ were chosen arbitrarily, results similar to those summarized in tables \ref{tabsp1}, \ref{tabsp2}, \ref{tabsp3}, \ref{tabnikkei1}, \ref{tabnikkei2}, \ref{tabnikkei3}, \ref{tabftse1}, \ref{tabftse2} and, \ref{tabftse3} are obtained using alternative parameter values.} Furthermore, for each combination of $N$ and $w$, we randomly sample 100 different sets of $N$ assets from the reference database and repeat the rolling window exercise to ensure that the results are not artifacts of a specific investment universe. Finally, since several of the strategies under comparison involve active management, we incorporate proportional transaction costs of 10 basis points (as supported by \cite{ledoit2025markowitz}) to ensure a realistic performance comparison.

\begin{remark}\label{rem3} 
As the empirical results demonstrate, transaction costs are negligible from a risk perspective, as they do not affect the variability of out-of-sample returns. Furthermore, the impact of these costs on risk is independent of the specific risk measure employed. Conversely, transaction costs naturally erode the average return of the portfolio, thereby reducing the numerator of the Sharpe ratio for active investment strategies.
\end{remark}

The performance of the various strategies is evaluated and compared using a comprehensive suite of metrics: average return (\textbf{a.r.}), standard deviation (\textbf{st.dev.}), the Sharpe ratio (\textbf{SR}), Value-at-Risk $(\textbf{VaR}_{1\%})$, Conditional Value-at-Risk $(\textbf{CVaR}_{1\%})$, maximum drawdown (\textbf{MDD}), skewness (\textbf{Sk}), and kurtosis (\textbf{K}) of the out-of-sample returns. In particular, \textbf{VaR} and \textbf{CVaR} are calculated at a $1\%$ significance level to specifically isolate and examine the behavior of the extreme left tail of the return distribution. This focus on the $1\%$ threshold ensures that the analysis captures the strategies' resilience during severe market dislocations, providing a rigorous assessment of their risk-mitigation properties.

The strategies under comparison are: 

\begin{itemize}
	\item the equally weighted portfolio, denoted by $\textbf{1/N}$, is the portfolio in which capital is allocated uniformly across each asset. This strategy serves as a quasi-passive benchmark. Indeed, the turnover of the $\textbf{1/N}$ portfolio is exceptionally low, arising solely from the need to rebalance the portfolio back to equal weights as asset prices fluctuate
	 \item the \textbf{risk reduction strategy}, hereafter referred to as $\textbf{RR}$, is defined as the investment approach inspired by the geometric risk-mitigation framework proposed in the paper. This strategy alternates between the equally weighted $\textbf{1/N}$ portfolio and a state with a reduced market exposure. Specifically, the strategy reduces exposure of $50\%$ when the highest risk scenario is detected, which occurs when the distance from the point representing matrix A to the origin is smaller than its distance to the vertex $[0,\ldots,0,1]$. In all other scenarios, where the risk is deemed affordable, the strategy remains fully invested in the $\textbf{1/N}$ benchmark portfolio.\footnote{The choice regarding the magnitude of exposure reduction is arbitrary and can be computed considering both the risk reduction and the return objectives.} When the exposure is reduced one can imagine that the correspondent wealth is allocated in liquidity; in this experiment no risk free returns is considered and the liquidity is accounted with a null return. This investment rule can be visualized by referring to Figure \ref{fig2} as the strategy that reduces exposure when the normalized spectrum falls within the regions identified by the red points, while maintaining a full investment position in all other cases.
	\item the so-called \textbf{random} strategy. This alternative benchmark investment strategy is constructed ex-post based on the $\textbf{RR}$ strategy's activity: first, we count the total number of days the $\textbf{RR}$ strategy signaled a reduction in exposure. We then build a strategy that randomly reduces exposure by the same amount for that same number of days. The implementation of this strategy serves two primary objectives. First, it demonstrates that a degree of risk reduction can be achieved trivially by simply reducing portfolio leverage, as noted in Remark \ref{rem1}. Second, and more importantly, it allows us to show that the risk reduction and performance metrics of the $\textbf{RR}$ strategy are superior to those of the random benchmark. This comparison underscores the significant information content of the geometric signal over a non-informative reduction in exposure.\footnote{Given the experimental design, the \textbf{random} strategy serves as a statistical test to determine whether the incremental risk reduction provided by the $\textbf{RR}$ strategy is significant.}
	\item Finally, we consider optimization-based approaches as additional benchmarks. These strategies are standard optimization frameworks where the optimal allocation is determined by solving a minimization problem. In this context, the objective is to minimize a specific risk measure, subject to the constraint that the optimal portfolio lies within the simplex (i.e., weights are non-negative and sum to unity). The risk measures employed are variance, VaR, and CVaR. Each investment strategy is identified by the risk measure used as the optimization objective and is named accordingly (e.g., \textbf{Min-var}, \textbf{Min-VaR}, or \textbf{Min-CVaR}).
\end{itemize}

\begin{remark}
Maximum drawdown is not typically used to calculate an optimal allocation due to the difficulty in forecasting path-dependent sequences of losses, see \cite{magdonismail2004}. Nevertheless, this unpredictability makes the metric of primary importance for evaluating the risk-reduction capacity of strategies. As the results of this application will demonstrate, this measure is not trivially sensitive to scale or leverage transformations; thus, its effective reduction can be interpreted as proof of a strategy's genuine risk-mitigation capability. Moreover, managing large drawdowns is critical for long-term survival, as deep sequences of losses are the principal cause of portfolio insolvency, see \cite{grossman1993optimal}. For the above reasons, the MDD is used as a metric for performance comparison evaluation while not used for the computation of an alternative optimal allocation.
\end{remark}

Historical parameter estimation has been implemented for the optimization-based techniques without applying any specialized estimation filters. This point is fundamental: as highlighted in Section \ref{lit}, a large portion of the literature devoted to standard optimization-based allocation issues focuses on the estimation of the optimization inputs. Naturally, if the estimation is accurate, implying high-quality forecasts, only relative issues arise from the implementation of an optimization approach. The central point is that while traditional methods require precise input forecasts to perform well, the approach proposed in this paper shifts the perspective entirely. Here, forecasting has the sole objective of determining whether the level of risk is acceptable or excessive. The gray zone between algebraic and numerical rank can be viewed as the empirical counterpart to the distinction between the concepts of risk and uncertainty, which are closely related but not equivalent. A fundamental distinction exists between risk and uncertainty, first formalized by \cite{knight1921risk}. \textit{Risk} refers to situations where potential outcomes are unknown but follow a measurable probability distribution, allowing for statistical optimization. In contrast, \textit{Knightian uncertainty} describes environments where the probability distribution itself is unknown or non-existent, rendering traditional probabilistic models insufficient and necessitating more robust, structural approaches to portfolio protection.

Before presenting the results of the comprehensive experiment across the three databases and for all combinations of $N$ and $w$, we report a single experiment to provide a graphical representation. This example serves to highlight the technical details involved in the practical application of the strategy. We select the S\&P500 database and set $N=15$ and $w=15$. While this choice is arbitrary, as will become clear, the results are highly robust to changes in the parameters $N$ and $w$, as well as to the use of different databases. To further simplify the interpretation and graphical representation of this example, the application is restricted to a sub-sample containing the last 240 observations, corresponding approximately to the final year of the database.

\begin{remark}\label{Nw}
The selection of the window length $w$ relative to the number of assets $N$ is strategic: larger values of $w$ compared to $N$ would \textit{artificially} ensure the algebraic full-column rank of the return matrix, potentially masking the \textbf{RR} strategy's ability to mitigate risk. To rigorously test the underlying theory, specifically the detection of risk arising from the loss of diversification opportunities via potential rank deficiency, the ideal configuration would be $w=N$. While this aspect is not the primary focus of the overall application, where in eight out of nine combinations $w>N$, it is noteworthy that compelling results are still achieved in these cases. Furthermore, as risk management is fundamentally a forecasting exercise, the estimation window length is a critical parameter. If $w$ is excessively long, the inclusion of stale data may cause the normalized spectrum to detect high-risk scenarios with a fatal delay. The fact that the proposed approach is designed to perform effectively with short estimation windows is of central importance. This stands in contrast to standard optimization-based techniques, which typically require long estimation windows, not only for parameter stability but also to mitigate the performance-eroding impact of transaction costs.
\end{remark} 

\begin{figure}[h!]
\centering
\includegraphics[width=0.99\linewidth]{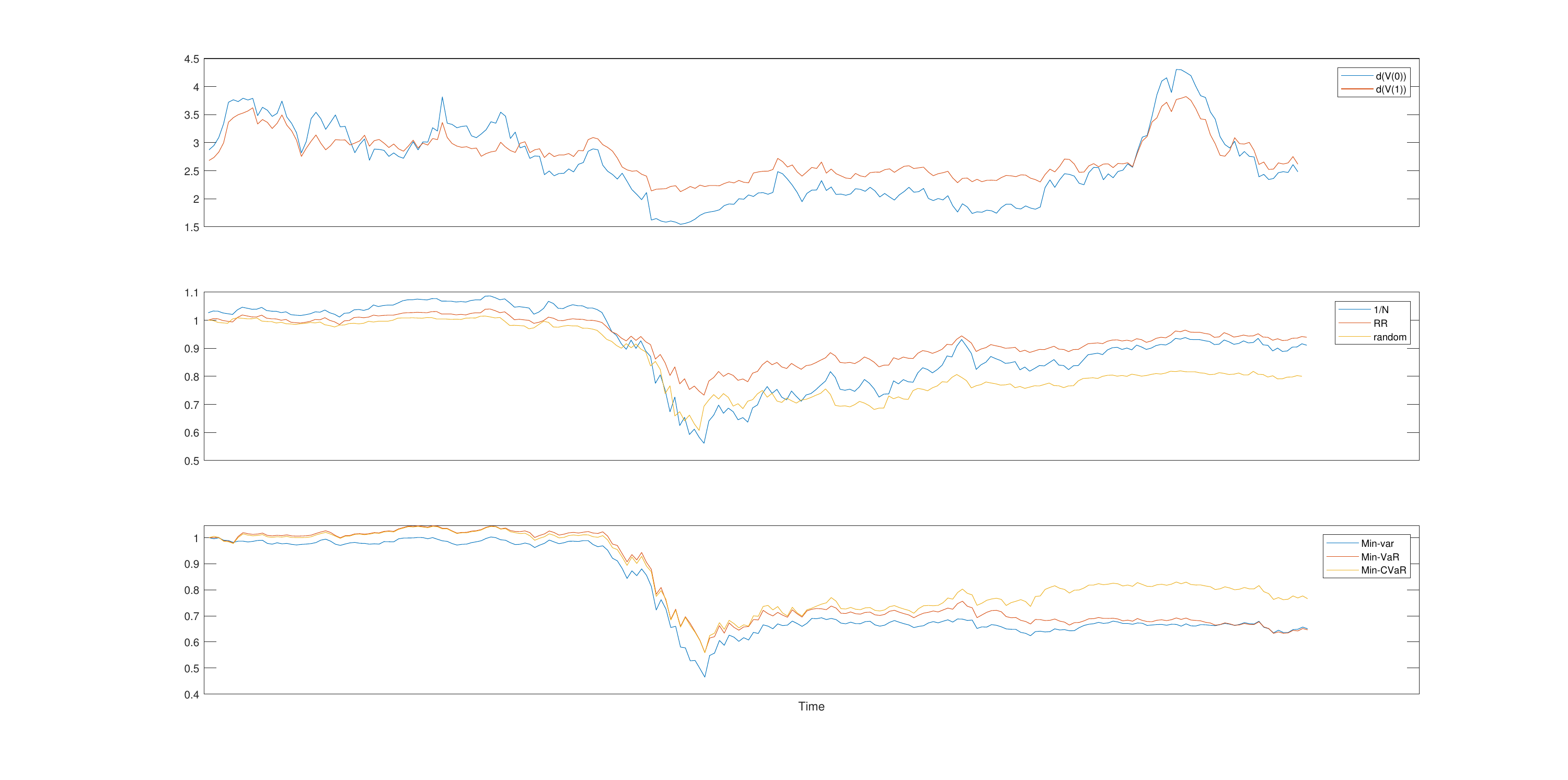}
\caption{Top plot: variation over time of the distances to the vertices on which the \textbf{RR} strategy is based; $d(V(0))$ is the distance to the origin and $d(V(1))$ is the distance to the vertex with $N-2$ null coordinates. Central plot: value over time of the portfolios $\textbf{1/N}$, \textbf{RR} and \textbf{random}. Bottom plot: value over time of the portfolios \textbf{Min-var}, \textbf{Min-VaR} and \textbf{Min-CVaR}.}\label{fig3}
\end{figure}

Figure \ref{fig3} contains three plots. The top graph reports the evolution over time of the distances from the hypercube vertices used to calculate trading signals; specifically, d(V(0)), represented by the blue line, is the distance to the origin, while d(V(1)), the red line, is the distance to the vertex with $N-2$ null coordinates. According to the strategy previously described, the \textbf{RR} strategy reduces exposure (leverage) when the blue line is below the red one, maintaining a fully invested position otherwise. An interesting observation is that the two lines do not cross continuously; rather, their relative positions remain stable for extended periods. In practice, this translates into a strategy that is simple to implement, as market exposure remains stable, reducing active management and, consequently, transaction costs. The central plot of Figure \ref{fig3} illustrates the evolution of portfolio values for three strategies: $\textbf{1/N}$, \textbf{RR}, and \textbf{random}. The bottom plot shows the values of the optimization-based approaches over the same period. These two representations have been separated to improve readability. The clearest evidence is the reduced volatility of the \textbf{RR} strategy and its effective timing, as it drastically mitigates the significant drawdown corresponding to the COVID-19 crisis.

\begin{figure}[!h]
\centering
\includegraphics[width=0.7\linewidth]{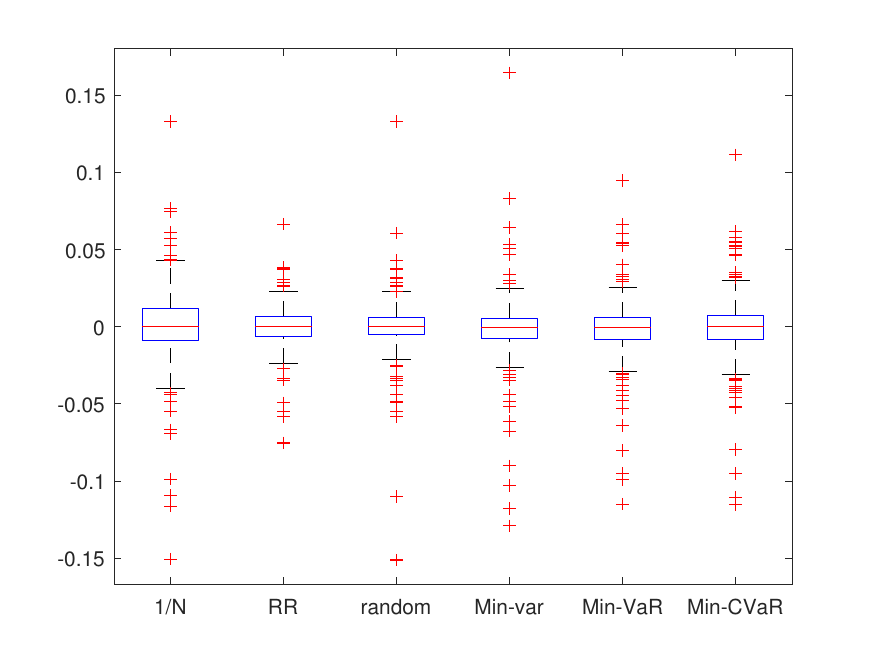}
\caption{Box Plot: comparison between the out-of-sample returns of the investment strategies under comparison.}\label{fig4}
\end{figure}

The effectiveness of the \textbf{RR} strategy in reducing risk relative to the other strategies is even more evident when comparing the out-of-sample returns through a box-plot representation (see Figure \ref{fig4}). All strategies appear very similar in terms of average return, as this value is typically very small and close to zero within a daily framework. Taking the $\textbf{1/N}$ as the reference benchmark, the primary evidence is the capacity of the \textbf{RR} strategy to strongly reduce the dispersion of out-of-sample returns while drastically mitigating the largest losses. Regarding the random strategy, it is interesting to observe a general capacity for reducing volatility due to leverage reduction; however, it is clear that random exposure reductions are unable to time the market and, consequently, avoid large losses. Optimization-based techniques show only slightly better return dispersion than the $\textbf{1/N}$ and appear far less effective than the \textbf{RR} strategy. Furthermore, optimization-based techniques seem only marginally able to prevent extreme losses.

The overall results of the experiment on the S\&P500 for all combinations of $w$ and $N$ are collected in Tables \ref{tabsp1}, \ref{tabsp2}, and \ref{tabsp3}. The results for the other databases are presented in Tables \ref{tabnikkei1}, \ref{tabnikkei2}, and \ref{tabnikkei3} for the Nikkei, and Tables \ref{tabftse1}, \ref{tabftse2}, and \ref{tabftse3} for the FTSE (see Appendix \ref{app}). For a given database and combination of $N$ and $w$, 100 rolling-window experiments were implemented across the entire sample for $N$ randomly selected assets; therefore, the summary tables contain the average value of these 100 experiments for each metric. Since the evidence is qualitatively similar across the different databases, the comments focus on the S\&P500, with the understanding that they can be analogously extended to the remaining datasets.

%%%%%%%%%%%%%%%%%%%%%%%%%%%%%%%%%%%%%%%%%%%%%%%%%%%%%%%%%%%%%%%%%%%%%%%%%%%%%%%%%%%%%%%%%%%%%%%%%%%%%%%%%%%%%%%%%
%%tabella sp500

\begin{table}[!h]
  \centering
  \caption{Comparison between the out-of-sample returns of the investment strategies for different values of $N = 5, 10, 20$ and  $w = 20$. Database: S\&P500.}
	\label{tabsp1}%
	\begin{adjustbox}{width=\textwidth}
    \begin{tabular}{|c|l|rrrrrr|}
    \toprule
    \multicolumn{1}{|r}{} &       & \multicolumn{6}{c|}{w = 20} \\
    \multicolumn{1}{|r}{} &       & \multicolumn{1}{c}{\textbf{1/N}} & \multicolumn{1}{c}{\textbf{RR}} & \multicolumn{1}{c}{\textbf{random}} & \multicolumn{1}{c}{\textbf{Min-var}} & \multicolumn{1}{c}{\textbf{Min-VaR}} & \multicolumn{1}{c|}{\textbf{Min-CVaR}} \\
    \midrule
    \multirow{8}[2]{*}{N = 5} & \textbf{a.r.} & 0.000308 & 0.000145 & 3.21E-05 & 0.00015 & 0.000104 & 9.98E-05 \\
          & \textbf{st. dev.} & 0.015605 & 0.009172 & 0.010321 & 0.013329 & 0.013797 & 0.014147 \\
          & \textbf{SR} & 0.0209 & 0.016775 & 0.00365 & 0.011638 & 0.008201 & 0.008257 \\
          & \textbf{VaR 1\%} & 0.045965 & 0.027964 & 0.0318 & 0.038936 & 0.040481 & 0.041767 \\
          & \textbf{CVaR 1\%} & 0.066546 & 0.037943 & 0.045896 & 0.057431 & 0.058427 & 0.060092 \\
          & \textbf{MDD} & 0.602217 & 0.433523 & 0.473687 & 0.558721 & 0.595913 & 0.601933 \\
          & \textbf{Sk } & -0.33302 & -0.38313 & -0.38723 & -0.36888 & -0.28994 & -0.31956 \\
          & \textbf{K } & 12.21057 & 10.19242 & 17.208 & 16.70028 & 13.44057 & 13.67598 \\
    \midrule
    \multirow{8}[2]{*}{N = 10} & \textbf{a.r.} & 0.000312 & 0.000157 & -3.19E-05 & 7.48E-05 & 7.99E-05 & 0.000133 \\
          & \textbf{st. dev.} & 0.014688 & 0.008961 & 0.011004 & 0.01191 & 0.012412 & 0.012637 \\
          & \textbf{SR} & 0.021306 & 0.017376 & -0.00296 & 0.006642 & 0.006497 & 0.010802 \\
          & \textbf{VaR 1\%} & 0.043613 & 0.027174 & 0.033211 & 0.035377 & 0.036561 & 0.037235 \\
          & \textbf{CVaR 1\%} & 0.065559 & 0.036598 & 0.051118 & 0.052797 & 0.054373 & 0.055563 \\
          & \textbf{MDD} & 0.596129 & 0.413421 & 0.541068 & 0.572247 & 0.617556 & 0.58633 \\
          & \textbf{Sk } & -0.53741 & -0.45498 & -0.66408 & -0.64449 & -0.53161 & -0.53066 \\
          & \textbf{K } & 15.09558 & 9.192014 & 21.10744 & 17.29822 & 14.68256 & 14.88402 \\
    \midrule
    \multirow{8}[2]{*}{N = 20} & \textbf{a.r.} & 0.000305 & 0.000164 & 1.22E-05 & 1.47E-05 & 4.25E-05 & 0.000101 \\
          & \textbf{st. dev.} & 0.013748 & 0.008575 & 0.010677 & 0.010444 & 0.01106 & 0.011397 \\
          & \textbf{SR} & 0.02233 & 0.019135 & 0.001161 & 0.001816 & 0.004205 & 0.009107 \\
          & \textbf{VaR 1\%} & 0.040463 & 0.026547 & 0.033172 & 0.031212 & 0.033083 & 0.033763 \\
          & \textbf{CVaR 1\%} & 0.061939 & 0.034927 & 0.049732 & 0.0461 & 0.04772 & 0.049215 \\
          & \textbf{MDD} & 0.567638 & 0.407322 & 0.535322 & 0.531703 & 0.544479 & 0.531254 \\
          & \textbf{Sk } & -0.5957 & -0.52195 & -0.63094 & -0.2425 & -0.17003 & -0.2818 \\
          & \textbf{K } & 15.40137 & 8.745777 & 20.4884 & 23.83934 & 20.38149 & 18.31367 \\
    \bottomrule
    \end{tabular}%
		\end{adjustbox}
  
\end{table}

\begin{table}[!h]
  \centering
  \caption{Comparison between the out-of-sample returns of the investment strategies for different values of $N = 5, 10, 20$ and  $w = 30$. Database: S\&P500.}
	\label{tabsp2}%
	\begin{adjustbox}{width=\textwidth}
    \begin{tabular}{|c|l|rrrrrr|}
    \toprule
    \multicolumn{1}{|r}{} &       & \multicolumn{6}{c|}{w = 30} \\
    \multicolumn{1}{|r}{} &       & \multicolumn{1}{c}{\textbf{1/N}} & \multicolumn{1}{c}{\textbf{RR}} & \multicolumn{1}{c}{\textbf{random}} & \multicolumn{1}{c}{\textbf{Min-var}} & \multicolumn{1}{c}{\textbf{Min-VaR}} & \multicolumn{1}{c|}{\textbf{Min-CVaR}} \\
    \midrule
    \multirow{8}[2]{*}{N = 5} & \textbf{a.r.} & 0.000299 & 0.000125 & 3.51E-05 & 0.000212 & 0.000186 & 0.000263 \\
          & \textbf{st. dev.} & 0.016842 & 0.00973 & 0.011033 & 0.013794 & 0.014554 & 0.014511 \\
          & \textbf{SR} & 0.01806 & 0.013143 & 0.003314 & 0.01578 & 0.013298 & 0.018774 \\
          & \textbf{VaR 1\%} & 0.050178 & 0.029322 & 0.033893 & 0.040783 & 0.043269 & 0.042406 \\
          & \textbf{CVaR 1\%} & 0.07459 & 0.041258 & 0.049911 & 0.060678 & 0.062623 & 0.06271 \\
          & \textbf{MDD} & 0.647939 & 0.460423 & 0.544158 & 0.577763 & 0.597168 & 0.558728 \\
          & \textbf{Sk } & -0.50541 & -0.49057 & -0.43661 & -0.3663 & -0.3023 & -0.26981 \\
          & \textbf{K } & 14.36097 & 11.42031 & 20.49103 & 16.59199 & 12.87858 & 17.86406 \\
    \midrule
    \multirow{8}[2]{*}{N = 10} & \textbf{a.r.} & 0.000309 & 0.000172 & -7.01E-06 & 0.000138 & 9.78E-05 & 0.000174 \\
          & \textbf{st. dev.} & 0.013467 & 0.008642 & 0.010475 & 0.010614 & 0.01147 & 0.011103 \\
          & \textbf{SR} & 0.023194 & 0.019998 & -0.00065 & 0.012799 & 0.008324 & 0.015457 \\
          & \textbf{VaR 1\%} & 0.039573 & 0.026554 & 0.031835 & 0.030559 & 0.033591 & 0.031209 \\
          & \textbf{CVaR 1\%} & 0.059931 & 0.035011 & 0.047506 & 0.046333 & 0.050311 & 0.04758 \\
          & \textbf{MDD} & 0.549 & 0.379931 & 0.488785 & 0.492876 & 0.544431 & 0.497177 \\
          & \textbf{Sk } & -0.49034 & -0.41517 & -0.35981 & -0.39253 & -0.50474 & -0.2634 \\
          & \textbf{K } & 15.30817 & 8.713233 & 19.69491 & 20.47969 & 16.1307 & 16.63061 \\
    \midrule
    \multirow{8}[2]{*}{N = 20} & \textbf{a.r.} & 0.000279 & 0.000151 & -4.06E-05 & 7.55E-05 & 8.19E-05 & 0.000107 \\
          & \textbf{st. dev.} & 0.014146 & 0.008601 & 0.011027 & 0.010272 & 0.011548 & 0.010983 \\
          & \textbf{SR} & 0.020026 & 0.017814 & -0.00354 & 0.007879 & 0.007571 & 0.010156 \\
          & \textbf{VaR 1\%} & 0.043309 & 0.026913 & 0.033638 & 0.029616 & 0.034971 & 0.032013 \\
          & \textbf{CVaR 1\%} & 0.06614 & 0.036119 & 0.052728 & 0.045747 & 0.05171 & 0.048604 \\
          & \textbf{MDD} & 0.583592 & 0.393896 & 0.547973 & 0.485377 & 0.546805 & 0.514234 \\
          & \textbf{Sk } & -0.69053 & -0.58757 & -0.51814 & -0.07019 & -0.56371 & -0.23682 \\
          & \textbf{K } & 17.25247 & 9.859038 & 23.34833 & 30.32789 & 17.36597 & 24.30235 \\
    \bottomrule
    \end{tabular}%
  \end{adjustbox}
\end{table}%

\begin{table}[!h]
  \centering
  \caption{Comparison between the out-of-sample returns of the investment strategies for different values of $N = 5, 10, 20$ and  $w = 40$. Database: S\&P500.}
	\label{tabsp3}%
	\begin{adjustbox}{width=\textwidth}
    \begin{tabular}{|c|l|rrrrrr|}
    \toprule
    \multicolumn{1}{|r}{} &       & \multicolumn{6}{c|}{w = 40} \\
    \multicolumn{1}{|r}{} &       & \multicolumn{1}{c}{\textbf{1/N}} & \multicolumn{1}{c}{\textbf{RR}} & \multicolumn{1}{c}{\textbf{random}} & \multicolumn{1}{c}{\textbf{Min-var}} & \multicolumn{1}{c}{\textbf{Min-VaR}} & \multicolumn{1}{c|}{\textbf{Min-CVaR}} \\
    \midrule
    \multirow{8}[2]{*}{N = 5} & \textbf{a.r.} & 0.000191 & 7.97E-05 & -7.01E-05 & 0.000138 & 0.000139 & 0.000125 \\
          & \textbf{st. dev.} & 0.017476 & 0.01008 & 0.011902 & 0.014264 & 0.014912 & 0.014933 \\
          & \textbf{SR} & 0.011595 & 0.008335 & -0.00569 & 0.010955 & 0.010159 & 0.009744 \\
          & \textbf{VaR 1\%} & 0.054391 & 0.031089 & 0.036865 & 0.042955 & 0.044739 & 0.044834 \\
          & \textbf{CVaR 1\%} & 0.079601 & 0.042896 & 0.055405 & 0.062785 & 0.065489 & 0.065142 \\
          & \textbf{MDD} & 0.706202 & 0.511409 & 0.597818 & 0.589453 & 0.621007 & 0.630665 \\
          & \textbf{Sk } & -0.56607 & -0.46462 & -0.48204 & -0.40935 & -0.32324 & -0.25887 \\
          & \textbf{K } & 16.74814 & 11.91402 & 25.0489 & 16.6094 & 15.22485 & 14.86031 \\
    \midrule
    \multirow{8}[2]{*}{N = 10} & \textbf{a.r.} & 0.000256 & 0.000124 & -9.44E-05 & 0.000134 & 0.000142 & 0.00017 \\
          & \textbf{st. dev.} & 0.014918 & 0.009158 & 0.011382 & 0.010966 & 0.011711 & 0.011622 \\
          & \textbf{SR} & 0.01767 & 0.014132 & -0.00803 & 0.012347 & 0.012173 & 0.014755 \\
          & \textbf{VaR 1\%} & 0.044563 & 0.028092 & 0.036069 & 0.031232 & 0.0342 & 0.033473 \\
          & \textbf{CVaR 1\%} & 0.06783 & 0.037805 & 0.053982 & 0.048418 & 0.051623 & 0.051278 \\
          & \textbf{MDD} & 0.600569 & 0.429548 & 0.57554 & 0.47241 & 0.50548 & 0.506823 \\
          & \textbf{Sk } & -0.51913 & -0.49148 & -0.76242 & -0.52277 & -0.36794 & -0.40063 \\
          & \textbf{K } & 17.14323 & 9.973636 & 22.98354 & 26.03392 & 22.22204 & 23.54282 \\
    \midrule
    \multirow{8}[2]{*}{N = 20} & \textbf{a.r.} & 0.000292 & 0.000154 & -2.89E-05 & 9.43E-05 & 0.000108 & 0.000152 \\
          & \textbf{st. dev.} & 0.014253 & 0.008633 & 0.010834 & 0.010618 & 0.011662 & 0.011479 \\
          & \textbf{SR} & 0.02057 & 0.017941 & -0.00253 & 0.009239 & 0.009654 & 0.013518 \\
          & \textbf{VaR 1\%} & 0.043239 & 0.026403 & 0.032997 & 0.031902 & 0.034325 & 0.034507 \\
          & \textbf{CVaR 1\%} & 0.065377 & 0.035855 & 0.051585 & 0.047802 & 0.052184 & 0.051714 \\
          & \textbf{MDD} & 0.593879 & 0.405429 & 0.524191 & 0.505304 & 0.541585 & 0.571822 \\
          & \textbf{Sk } & -0.63158 & -0.55019 & -0.79259 & -0.50936 & -0.58559 & -0.61505 \\
          & \textbf{K } & 15.83105 & 9.565252 & 24.24013 & 18.99702 & 16.75699 & 17.45016 \\
    \bottomrule
    \end{tabular}%
  \end{adjustbox}
\end{table}%

%%%%%%%%%%%%%%%%%%%%%%%%%%%%%

%%%%%%%%%%%%%%%%%%%%%%%%%%%%%%%%%%%%%%%%%%%%%%%%%%%%%%%%%%%%%%%%%%%%%%%%%%%%%%%%%%%%%%%%%%%%%%%%%%%%%%%%%%%%%%%%%
 Focusing first on the risk perspective, the \textbf{RR} strategy is associated with the lowest risk across all combinations of $w$ and $N$, regardless of the risk measure considered. Moreover, the risk reduction is consistently greater than that obtained by the \textbf{random} strategy, supporting the fact that this reduction is not merely a trivial, deterministic byproduct of random leverage reduction. In contrast, the risk reduction obtained by implementing risk-minimization approaches is not only less effective than that of the \textbf{RR} strategy but is often comparable to the results of the random strategy. This finding is particularly concerning, as it suggests a significant random component in the performance of traditional optimization results. Let us emphasize that the risk reduction achieved through standard optimization-based techniques is entirely attributable to the specific allocation of wealth among the assets. In contrast, the risk reduction observed in the \textbf{RR} strategy is primarily driven by the timely and strategic reduction of portfolio leverage. A specific comment is necessary regarding the maximum drawdown (\textbf{MDD}). The \textbf{RR} strategy demonstrates a significant capacity to reduce \textbf{MDD} in every combination of $w$ and $N$, a reduction that the random strategy fails to replicate. The explanation for this is fundamental to understanding the risk-mitigation power of the proposed strategy: while randomly reducing exposure has a non-negligible effect on standard risk measures that account for variability, this ``trick'' does not effectively mitigate sequences of losses. This empirical evidence provides a strong argument supporting the risk-detection capacity of our proposal. Standard optimization-based allocations show a lower capacity to reduce \textbf{MDD} than the \textbf{RR} strategy, though they generally outperform the random strategy.
Regarding the higher moments of the return distribution, the \textbf{RR} strategy is the only strategy that in many cases improves the distribution's shape. Specifically, it tends to increase (make less negative) the skewness and significantly reduce the kurtosis compared to the $\textbf{1/N}$ portfolio. This indicates that the strategy successfully prunes the ``fat tails'' of the return distribution, further validating its effectiveness in avoiding extreme negative realizations. On the contrary, all the other strategies generally increase kurtosis while reducing skewness. This shift in the higher moments suggests that standard approaches may inadvertently heighten the probability of extreme negative outcomes during periods of market stress. By contrast, the \textbf{RR} strategy's ability to move toward a more symmetric and thinner-tailed distribution highlights its structural advantage in risk mitigation.

Regarding the return side, while not the primary focus of this research, our findings align with a well-known result in the literature: most active portfolio strategies exhibit a Sharpe ratio lower than that of the \textbf{1/N} benchmark, as documented by \cite{DeMig}. This contraction is entirely attributable to transaction costs that erode returns, given that all active strategies evaluated here actually achieve lower risk profiles than the $\textbf{1/N}$ portfolio. The fact that the return of the $\textbf{1/N}$ portfolio could be associated to a higher level of risk is a known and studied topic in the literature, see e.g. \cite{hwang2018naive}.
Notably, the \textbf{RR} strategy consistently outperforms other active strategies in terms of the Sharpe ratio (\textbf{SR}). Furthermore, its \textbf{SR} decay relative to the $\textbf{1/N}$ benchmark is minimal. This resilience stems from both the magnitude of its risk reduction and its operational efficiency; as shown in Figure \ref{fig3}, alarm signals for reduced diversification opportunities typically arrive in sequences. This stability makes the strategy simple to implement in practice, significantly limiting the impact of transaction costs compared to other active methods. While standard optimization approaches frequently oscillate between optimal allocations from one period to the next, creating instability and high turnover, the \textbf{RR} strategy maintains exposure reductions over continuous periods.

If a practitioner's objective extends beyond risk mitigation to include return enhancement, the strategy could be refined by calibrating the specific proportion of exposure reduction. Another possibility would be to scale the level of deleveraging proportionally with the degree of potential rank loss identified by the proximity to the hypercube vertices. However, such refinements are beyond the scope of the current research, which is designed exclusively to introduce a new perspective on risk detection.

One may think that the results obtained are a consequence of the short estimation window, which is only slightly larger than the portfolio size $N$. A significant branch of the literature supports the idea that a long estimation window is necessary, particularly for the estimation of the covariance matrix, which contains $\frac{N(N+1)}{2}$ distinct parameters (see, among others, 
\cite{ledoit2004well}, \cite{bickel2008regularized}, \cite{ledoit2017nonlinear}, and \cite{denard2021factor}). To investigate the role of the estimation window length $w$ in the present application, we implemented the experiment on the S\&P500 database using $N=10$ and $w=240$, representing approximately one year of daily data. The results for this experiment are summarized in Table \ref{tabwlong}.

%%%%%%%%%%%%%%%%%%%%%%%%%%%%%%%%%%%%%%%%%%%%%%%%%%%%%%%%%%%%%%%%%%%%%%%%%%%%%%%%%%%%%%%%%%%%%%%%%%%%%%%%%%%%%%%%%%%%%%%%%
%%%%%tab sp finestra lunga%%%%%%%%%%%%%%%%%%%%%%%%%%%%%%%%%%%%%%%%%%%%%%%%%%%%%%%%%%%%%%%%%%%%%%%%%%%%%%%%%%%%%%%%%%%%%%%
\begin{table}[!h]
  \centering
  \caption{Comparison between the out-of-sample returns of the investment strategies for $N = 10$ and $w = 240$. Database: S\&P500.}
	\label{tabwlong}
	\begin{adjustbox}{width=\textwidth}
    \begin{tabular}{|c|l|rrrrrr|}
    \toprule
    \multicolumn{1}{|r}{} &       & \multicolumn{6}{c|}{w = 240} \\
    \multicolumn{1}{|r}{} &       & \multicolumn{1}{c}{\textbf{1/N}} & \multicolumn{1}{c}{\textbf{RR}} & \multicolumn{1}{c}{\textbf{random}} & \multicolumn{1}{c}{\textbf{Min-var}} & \multicolumn{1}{c}{\textbf{Min-VaR}} & \multicolumn{1}{c|}{\textbf{Min-CVaR}} \\
    \midrule
    \multirow{8}[2]{*}{N = 10} & \textbf{a.r.} & 0.000271 & 0.000136 & -3.99E-05 & 0.000188 & 0.000137 & 0.000165 \\
          & \textbf{st. dev.} & 0.014675 & 0.009141 & 0.010574 & 0.011264 & 0.012607 & 0.011506 \\
          & \textbf{SR} & 0.018897 & 0.015603 & -0.00346 & 0.017242 & 0.011069 & 0.01483 \\
          & \textbf{VaR 1\%} & 0.043832 & 0.028452 & 0.032927 & 0.033328 & 0.037755 & 0.034412 \\
          & \textbf{CVaR 1\%} & 0.066356 & 0.039841 & 0.049403 & 0.051155 & 0.0576 & 0.052048 \\
          & \textbf{MDD} & 0.600705 & 0.435152 & 0.515869 & 0.503095 & 0.572992 & 0.526213 \\
          & \textbf{Sk } & -0.57959 & -0.64028 & -0.47148 & -0.63243 & -0.64883 & -0.64581 \\
          & \textbf{K } & 15.60355 & 12.40128 & 20.41001 & 20.13859 & 18.52564 & 18.43927 \\
    \bottomrule
    \end{tabular}%
		\end{adjustbox}
\end{table}%
%%%%%%%%%%%%%%%%%%%%%%%%%%%%%%%%%%%%%%%%%%%%%%%%%%%%%%%%%%%%%%%%%%%%%%%%%%%%%%%%%%%%%%%%%%%%%%%%%%%%%%%%%%%%%%%%%%%%%%%%%

Looking at Table \ref{tabwlong}, several interesting findings can be highlighted. First, even with a longer estimation window, the risk-reduction capacity of the optimization-based approaches remains less effective than that of the \textbf{RR} strategy. Moreover, the risk reduction achieved with $w=240$ is not significantly higher than that obtained with a shorter window. This phenomenon is a result of the well-known trade-off between the necessity of a long estimation window for estimating a large number of parameters and the drawback of using stale data, which has a limited capacity to forecast current market fluctuations. The only tangible advantage of using a longer estimation window is the increased parameter stability, which significantly improves the practical implementation of the strategy by making optimized portfolios more stable and reducing the impact of transaction costs. This is evident from the increased Sharpe ratios of the optimization-based strategies. What is particularly noteworthy is that the \textbf{RR} strategy is only marginally sensitive to the length of $w$. It remains the superior strategy in terms of risk reduction, while the impact on its Sharpe ratio is limited because its allocation is already stable even with a short $w$. Consequently, the influence of transaction costs does not vary significantly with the window length for the \textbf{RR} approach.

\begin{remark}\label{remul}
In academic research, risk reduction is usually associated with a diminished performance of investment strategies, as documented by \cite{ang2006cross}, \cite{asness2020betting}, and \cite{baker2011benchmarks}. Practitioners and investors often hold a more radical perspective: they contend that with perfect timing, it should be possible to prevent losses entirely. If one could avoid losses through flawless timing, average returns would increase while risk decreases, creating a simultaneous positive effect on both components of the Sharpe ratio. While such reasoning may be idealistic, a practitioner typically expects a risk-minimization model to reduce exposure during turbulent market periods. This shift rarely occurs when using optimization approaches that base diversification on the computation of co-risk (correlation). In practical terms, a practitioner would expect a model to suggest reducing market exposure during a major event like the 2008 financial crisis. To our knowledge, no standard allocation models behave in this manner. Even the Markowitz model, when a risk-free asset is included, often fails to shift the allocation toward liquidity during financial crises, likely due to fundamental forecasting limitations. In contrast, the \textbf{RR} strategy proposed in this paper is specifically designed to address this need, modulating portfolio leverage according to the detected level of risk. A final provocation of this paper concerns the true effectiveness of asset allocation for risk reduction. The empirical results presented here suggest that managing and timing the leverage exposure of a portfolio is more effective at mitigating risk than attempting to determine an ``optimal'' allocation among assets. This is because traditional allocation requires the accurate forecasting of future correlations, a task that remains notoriously difficult during the very periods of market stress when risk management is most critical.
This last aspect is more closely aligned with the uncertainty inherent in financial markets rather than their measurable riskiness. While risk can be managed through the precise calibration of asset weights based on past correlations, uncertainty involves the structural breakdown of those very relationships.
\end{remark}   

As highlighted in Remark \ref{remul}, this research suggests that the allocation itself is of secondary importance for risk reduction compared to the timing of changes in portfolio leverage (exposure). To test this, an ad hoc experiment was constructed. Using the S\&P500 database with $N=10$ and $w=20$, we modified the \textbf{RR} strategy by changing the benchmark portfolio. Specifically, while the exposure follows the same rule previously described, the underlying benchmark portfolio is selected by uniformly extracting a random point from the simplex in $\mathbb{R}^N$. Furthermore, since each experiment for a given $N$ and $w$ is repeated 100 times using randomly selected assets, this version of the experiment assigns a new, randomly chosen reference benchmark to each of the selected investment universes. This procedure is designed to stress the robustness of our findings as much as possible, effectively neutralizing any potential bias related to the specific choice of assets or the benchmark allocation. The results, summarized in Table \ref{tabrand}, demonstrate that the risk-mitigation properties of the \textbf{RR} signal remain dominant regardless of the specific internal weighting of the portfolio.

%%%%%%%%%%%%%%%%%%%%%%%%%%%%%%%%%%%%%%%%%%%%%%%%%%%%%%%%%%%%%%%%%%%%%%%%%%%%%%%%%%%%%%%%%%%%%%%%%%%%%%%%%%%%%%%%%%%%%%%%%%%%%%%%
%%%%%%%%%sp random benchmark%%%%%%%%%%%%%%%%%%%%%%%%%%%%%%%%%%%%%%%%%%%%%%%%%%%%%%%%%%%%%%%%%%%%%%%%%%%%%%%%%%%%%%%%%%%%%%%%%%%%

\begin{table}[!h]
  \centering
  \caption{Comparison between the out-of-sample returns of the investment strategies for $N = 10$ and $w = 20$, database: S\&P500, random select benchmark (rand bench).}
    \begin{tabular}{|c|l|rrr|}
    \toprule
    \multicolumn{1}{|r}{} &       & \multicolumn{3}{c|}{w = 20} \\
    \multicolumn{1}{|r}{} &       & \multicolumn{1}{c}{\textbf{rand bench}} & \multicolumn{1}{c}{\textbf{RR}} & \multicolumn{1}{c|}{\textbf{random}} \\
    \midrule
    \multirow{8}[2]{*}{N = 10} & \textbf{a.r.} & 0.000245 & 0.000117 & -8.36E-05 \\
          & \textbf{st. dev.} & 0.015459 & 0.00954 & 0.010989 \\
          & \textbf{SR} & 0.016053 & 0.012831 & -0.00732 \\
          & \textbf{VaR 1\%} & 0.046083 & 0.029169 & 0.033776 \\
          & \textbf{CVaR 1\%} & 0.069641 & 0.041458 & 0.052227 \\
          & \textbf{MDD} & 0.6127 & 0.446497 & 0.557495 \\
          & \textbf{Sk } & -0.5532 & -0.61068 & -0.80347 \\
          & \textbf{K } & 16.85982 & 15.25423 & 25.53661 \\
    \bottomrule
    \end{tabular}%
  \label{tabrand}%
\end{table}%

%%%%%%%%%%%%%%%%%%%%%%%%%%%%%%%%%%%%%%%%%%%%%%%%%%%%%%%%%%%%%%%%%
This comparison focuses exclusively on the random benchmark and the corresponding RR strategy, as the inclusion of optimization-based approaches would be irrelevant to the objectives of this experiment. The results are analogous to those obtained in the previous applications: the \textbf{RR} strategy, even when implemented by investing in a random portfolio of $N$ randomly selected assets, significantly reduces risk at the cost of a very limited reduction in performance. This reinforces the core thesis of the research: the structural signal extracted from the return matrix's rank is a powerful indicator of instability that provides effective protection regardless of the underlying asset allocation. By demonstrating that the \textbf{RR} mechanism adds value to a purely stochastic (random) benchmark, we isolate the timing of leverage as the primary driver of risk mitigation, rather than the sophistication of the weighting scheme itself.

While remaining generally skeptical of overly complex refinements, due to the risk of ad hoc solutions that could diminish the strategy's robust forecasting performance, we offer a suggestion on how to potentially enhance the already strong results of the application. A deeper analysis of this specific aspect is left for future research. To explore these potential improvements, we slightly modify the \textbf{RR} strategy into a version we call the \textbf{enhanced RR}. In this version, the portfolio leverage toggles between 0 and 1 based on identified risk scenarios: the portfolio remains fully invested in the $\textbf{1/N}$ benchmark when the risk level is deemed acceptable and shifts $100\%$ into liquidity when the risk is considered unacceptable. This experiment is implemented on the S\&P500 database with $N=10$ and $w=20$. The risk signal in this version is derived using a more aggregate measure. Instead of comparing the distances to the two vertices associated with the highest rank loss individually, we compute the
arithmetic mean of the two smallest singular values and compare it to the threshold $\frac{1}{N-1}$. If the arithmetic mean is larger  
than this threshold, the risk is considered acceptable and the portfolio remains fully invested; conversely, if the mean falls below the threshold, the portfolio shifts entirely into liquidity.

%%%%%%%%%%%%%%%%%%%%%%%%%%%%%%%%%%%%%%%%%%%%%%%%%%%%%%%%%%%%%%%%%%%%%%%%%%%%%%%%%%%%%%%%%%%%%%%%%%%%%%%%%%%%%%%%%%%%%%%%%%%%%%%%
%%%%%% tab over performance%%%%%%%%%%%%%%%%%%%%%%%%%%%%%%%%%%%%%%%%%%%%%%%%%%%%%%%%%%%%%%%%%%%%%%%%%%%%%%%%%%%%%%%%%%%%%%%%%%%%%

\begin{table}[!h]
  \centering
  \caption{The out-of-sample returns of the \textbf{enhanced RR} strategy for $N = 20$ and $w = 40$, database: S\&P500.}
    \begin{tabular}{|c|l|rrr|}
    \toprule
    \multicolumn{1}{|r}{} &       & \multicolumn{3}{c|}{w = 40} \\
    \multicolumn{1}{|r}{} &       & \multicolumn{1}{c}{\textbf{1/N}} & \multicolumn{1}{c}{\textbf{RR}} & \multicolumn{1}{c|}{\textbf{random}} \\
    \midrule
    \multirow{8}[2]{*}{N = 20} & \textbf{a.r.} & 0.000282 & 0.000205 & -0.00011 \\
          & \textbf{st. dev.} & 0.014227 & 0.008726 & 0.012539 \\
          & \textbf{SR} & 0.02008 & 0.023441 & -0.009 \\
          & \textbf{VaR 1\%} & 0.043145 & 0.027185 & 0.038832 \\
          & \textbf{CVaR 1\%} & 0.065735 & 0.035281 & 0.060667 \\
          & \textbf{MDD} & 0.596593 & 0.314883 & 0.641403 \\
          & \textbf{Sk } & -0.62643 & -0.55963 & -0.66975 \\
          & \textbf{K } & 16.18966 & 10.07746 & 20.63544 \\
    \bottomrule
    \end{tabular}%
  \label{tabextraperf}%
\end{table}%

%%%%%%%%%%%%%%%%%%%%%%%%%%%%%%%%%%%%%%%%%%%%%%%%%%%%%%%%%%%%%%%%%%%%%%%%%%%%%%%%%%%%%%%%%%%%%%%%%%%%%%%%%%%%%%%%%%%%%%%%%%%%%%%%
This experiment highlights several critical aspects of the strategy's evolution. First, one may choose to work directly with the singular values rather than the distances to the vertices of the hypercube. However, this shift necessitates evaluating the magnitude of the singular values and the selection of an appropriate threshold, reintroducing the complex challenges associated with determining numerical rank and its related concepts. Furthermore, the magnitude of the singular values and the thresholds used to evaluate their significance are strictly dependent on the relative sizes of $w$ and $N$. This relationship is a cornerstone of Random Matrix Theory (RMT), particularly when calculating the probability distribution of the eigenvalues of a correlation matrix \cite{mehta2004random, laloux1999random}. The results of the experiment, summarized in Table \ref{tabextraperf}, show that the \textbf{enhanced RR} strategy maintains its robust risk-reduction capabilities while significantly increasing the Sharpe ratio. These findings suggest the possibility of achieving what practitioners ultimately desire: substantial risk mitigation without penalizing performance. Nevertheless, the implementation of this strategy requires the development of a generalized criterion for identifying risky scenarios that simultaneously accounts for $w$, $N$, the threshold level, and the number of singular values considered. This multidimensional optimization remains an open question for further research.

\section{Conclusions}\label{concl}

In the presence of a tornado watch, one is rarely interested in calculating the precise probability that a tornado will strike their home at a specific hour; the rational response is simply to seek shelter in the basement. Following this logic, this research argues that rather than attempting the nearly impossible task of precisely measuring risk in dynamic markets, an investor should focus on identifying high-risk scenarios and protecting against extreme losses by reducing exposure when the ``storm'' is detected. Accordingly, we have proposed an alternative approach to risk detection that moves away from traditional, probability-based risk measures. This approach relies on a novel intuition linking the geometry of the return space to systemic risk, where risk is represented by the potential loss of rank in the return matrix. The financial interpretation is immediate: a loss of rank signifies a collapse in the dimensionality of the return space, which directly corresponds to a reduction in available diversification opportunities. By analyzing the normalized spectrum of the return matrix, risk scenarios are defined based on the distance to the vertices of a unitary hypercube, each representing a rank-deficient matrix with varying degrees of rank loss. The effectiveness of this approach was rigorously tested through comprehensive experiments on real data, comparing out-of-sample returns across several investment strategies. The key results of the application can be summarized as follows. The \textbf{RR} strategy, reducing exposure in correspondence with high risk scenarios, demonstrates a superior capacity for risk reduction compared to both standard optimization-based techniques and simple benchmark strategies (such as $\textbf{1/N}$). Across all tested databases (S\&P500, Nikkei, and FTSE), the strategy significantly reduced Maximum Drawdown and CVaR. The findings suggest that the timing of leverage reduction is fundamentally more important for capital preservation than the specific internal weighting of the assets. The strategy remains effective even with short estimation windows and random asset selections, proving that the geometric signal of rank loss is a robust indicator of market instability. Ultimately, by modulating exposure according to the detected level of geometric risk, the \textbf{RR} strategy offers a practical and resilient framework for navigating periods of high uncertainty where traditional diversification models typically fail.

\section*{Acknowledgments} The author would like to thank his family for the support.

\section*{Declarations}
\noindent
{\bf Conflict of interest.}
The author has no conflicts of interest to declare that are relevant to the content of this article.

\noindent 
{\bf Funding.}
No funding was received for this study.

\noindent
{\bf Data availability.}
Data are available upon reasonable request to the author.

%%%%%%%%%%%%%%%%%%%%%%%%%%%%%%%%%%%%%%%%%%%%%%%%%%%%%%%%%%%%%%%%%%%%%%%%%%%%%%%%%%%%%%%%%%%%%%%%%%%%%%%%%%%%%%%%%%%%%%%%%%%%%%%%%%
% --- Bibliography ---
\bibliographystyle{plainnat} % Or the style required by your target journal
\bibliography{bibliografia1}    % Points to your references.bib file

@article{maggi2020proper,
  title={Proper measures of connectedness},
  author={Maggi, Mario and Torrente, Maria-Laura and Uberti, Pierpaolo},
  journal={Annals of Finance},
  volume={16},
  number={4},
  pages={547--571},
  year={2020},
  publisher={Springer}
}

@article{DeMig,
  title={Optimal versus naive diversification: How inefficient is the 1/N portfolio strategy?},
  author={DeMiguel, Victor and Garlappi, Lorenzo and Uppal, Raman},
  journal={The review of Financial studies},
  volume={22},
  number={5},
  pages={1915--1953},
  year={2009},
  publisher={Oxford University Press}
}

@article{kim,
  title={Recent developments in robust portfolios with a worst-case approach},
  author={Kim, Jang Ho and Kim, Woo Chang and Fabozzi, Frank J},
  journal={Journal of Optimization Theory and Applications},
  volume={161},
  pages={103--121},
  year={2014},
  publisher={Springer}
}

@article{artzner1999coherent,
  title={Coherent measures of risk},
  author={Artzner, Philippe and Delbaen, Freddy and Eber, Jean-Marc and Heath, David},
  journal={Mathematical finance},
  volume={9},
  number={3},
  pages={203--228},
  year={1999},
  publisher={Wiley Online Library}
}

@article{follmer2002convex,
  title={Convex measures of risk and trading constraints},
  author={F{\"o}llmer, Hans and Schied, Alexander},
  journal={Finance and stochastics},
  volume={6},
  number={4},
  pages={429--447},
  year={2002},
  publisher={Springer}
}

@article{rockafellar2006generalized,
  title={Generalized deviations in risk analysis},
  author={Rockafellar, R Tyrrell and Uryasev, Stan and Zabarankin, Michael},
  journal={Finance and Stochastics},
  volume={10},
  number={1},
  pages={51--74},
  year={2006},
  publisher={Springer}
}

@incollection{acciaio2011dynamic,
  title={Dynamic risk measures},
  author={Acciaio, Beatrice and Penner, Irina},
  booktitle={Advanced mathematical methods for finance},
  pages={1--34},
  year={2011},
  publisher={Springer}
}

@article{M,
author = {Markowitz, Harry},
title = {PORTFOLIO SELECTION*},
journal = {The Journal of Finance},
volume = {7},
number = {1},
pages = {77-91},
year = {1952}
}

@article{mousavi2014risk,
  title={Risk, uncertainty, and heuristics},
  author={Mousavi, Shabnam and Gigerenzer, Gerd},
  journal={Journal of Business Research},
  volume={67},
  number={8},
  pages={1671--1678},
  year={2014},
  publisher={Elsevier}
}

@article{malladi2017equal,
  title={Equal-weighted strategy: Why it outperforms value-weighted strategies? Theory and evidence},
  author={Malladi, Rama and Fabozzi, Frank J},
  journal={Journal of Asset Management},
  volume={18},
  number={3},
  pages={188--208},
  year={2017},
  publisher={Springer}
}

@article{zakamulin2017superiority,
  title={Superiority of optimized portfolios to naive diversification: Fact or fiction?},
  author={Zakamulin, Valeriy},
  journal={Finance Research Letters},
  volume={22},
  pages={122--128},
  year={2017},
  publisher={Elsevier}
}

@article{bartesaghi2025global,
  title={Global balance and systemic risk in financial correlation networks},
  author={Bartesaghi, Paolo and Diaz-Diaz, Fernando and Grassi, Rosanna and Uberti, Pierpaolo},
  journal={Physica A: Statistical Mechanics and its Applications},
  pages={130698},
  year={2025},
  publisher={Elsevier}
}

@article{pastorino2024empirical,
  title={An empirical comparison of correlation-based systemic risk measures},
  author={Pastorino, Caterina and Uberti, Pierpaolo},
  journal={Quality \& Quantity},
  volume={58},
  number={3},
  pages={2289--2314},
  year={2024},
  publisher={Springer}
}

@article{figini2020market,
  title={The market rank indicator to detect financial distress},
  author={Figini, Silvia and Maggi, Mario and Uberti, Pierpaolo},
  journal={Econometrics and Statistics},
  volume={14},
  pages={63--73},
  year={2020},
  publisher={Elsevier}
}

@article{demiguel2009generalized,
  title={A generalized approach to portfolio optimization: Improving performance by constraining portfolio norms},
  author={DeMiguel, Victor and Garlappi, Lorenzo and Nogales, Francisco J and Uppal, Raman},
  journal={Management science},
  volume={55},
  number={5},
  pages={798--812},
  year={2009},
  publisher={INFORMS}
}

@article{allen2019defense,
  title={In defense of portfolio optimization: What if we can forecast?},
  author={Allen, David and Lizieri, Colin and Satchell, Stephen},
  journal={Financial Analysts Journal},
  volume={75},
  number={3},
  pages={20--38},
  year={2019},
  publisher={Taylor \& Francis}
}

@article{kritzman2010defense,
  title={In defense of optimization: the fallacy of 1/N},
  author={Kritzman, Mark and Page, S{\'e}bastien and Turkington, David},
  journal={Financial Analysts Journal},
  volume={66},
  number={2},
  pages={31--39},
  year={2010},
  publisher={Taylor \& Francis}
}

@article{ghahtarani2022robust,
  title={Robust portfolio selection problems: a comprehensive review},
  author={Ghahtarani, Alireza and Saif, Ahmed and Ghasemi, Alireza},
  journal={Operational Research},
  volume={22},
  number={4},
  pages={3203--3264},
  year={2022},
  publisher={Springer}
}

@article{gelmini2024equally,
  title={The equally weighted portfolio still remains a challenging benchmark},
  author={Gelmini, Matteo and Uberti, Pierpaolo},
  journal={International Economics},
  volume={179},
  pages={100525},
  year={2024},
  publisher={Elsevier}
}

@article{frittelli2002putting,
  title={Putting order in risk measures},
  author={Frittelli, Marco and Gianin, Emanuela Rosazza},
  journal={Journal of Banking \& Finance},
  volume={26},
  number={7},
  pages={1473--1486},
  year={2002},
  publisher={Elsevier}
}

@article{acerbi2002spectral,
  title={Spectral measures of risk: A coherent representation of subjective risk aversion},
  author={Acerbi, Carlo},
  journal={Journal of banking \& finance},
  volume={26},
  number={7},
  pages={1505--1518},
  year={2002},
  publisher={Elsevier}
}

@article{ledoit2025markowitz,
  title={Markowitz portfolios under transaction costs},
  author={Ledoit, Olivier and Wolf, Michael},
  journal={The quarterly review of economics and finance},
  volume={100},
  pages={101962},
  year={2025},
  publisher={Elsevier}
}

@article{danielsson2008blame,
  title={Blame the models},
  author={Danielsson, Jon},
  journal={Journal of Financial Stability},
  volume={4},
  number={4},
  pages={321--328},
  year={2008},
  publisher={Elsevier}
}

@article{danielsson2016model,
  title={Model risk of risk models},
  author={Danielsson, Jon and James, Kevin R and Valenzuela, Marcela and Zer, Ilknur},
  journal={Journal of Financial Stability},
  volume={23},
  pages={79--91},
  year={2016},
  publisher={Elsevier}
}

@article{daniels2026economic,
  title={Economic conditions and portfolio tail risk: A probability-weighted simulation approach},
  author={Jiao, Lei and others},
  journal={Journal of Empirical Finance},
  pages={101715},
  year={2026},
  publisher={Elsevier}
}

@techreport{oecd2026catastrophic,
  title={Financial Protection Against Catastrophic Risks: 2026 Global Assessment},
  author={{OECD}},
  year={2026},
  institution={Organisation for Economic Co-operation and Development},
  type={Report},
  url={https://www.oecd.org/finance/catastrophic-risks}
}

@article{meucci2009managing,
  title={Managing diversification},
  author={Meucci, Attilio},
  journal={Risk},
  volume={22},
  number={5},
  pages={74--79},
  year={2009}
}

@article{meucci2013measuring,
  title={Measuring portfolio diversification based on optimized uncorrelated factors},
  author={Meucci, Attilio and Santangelo, Alberto and Deguest, Romain},
  journal={The Journal of Portfolio Management},
  volume={40},
  number={1},
  pages={70--82},
  year={2013},
  publisher={Institutional Investor Journals Umbrella}
}

@article{bera2008optimal,
  title={Optimal portfolio diversification using the maximum entropy principle},
  author={Bera, Anil K and Park, Sung Y},
  journal={Econometric Reviews},
  volume={27},
  number={4-6},
  pages={484--512},
  year={2008},
  publisher={Taylor \& Francis}
}

@techreport{imf2026diversification,
  title={Stock-Bond Diversification Offers Less Protection From Market Selloffs},
  author={Adrian, Tobias and Kramer, Johannes and Malik, Sheheryar},
  year={2026},
  institution={International Monetary Fund},
  type={Global Financial Stability Analysis},
  url={https://www.imf.org/en/blogs/articles/2026/02/18/stock-bond-diversification-offers-less-protection-from-market-selloffs}
}

@book{ecb2025connectedness,
  title={Investor sentiment and dynamic connectedness in European markets: insights from the covid-19 and Russia-Ukraine conflict},
  author={Bouteska, Ahmed and Buchetti, Bruno and Harasheh, Murad and Santoni, Alessandro},
  number={3050},
  year={2025},
  publisher={ECB Working Paper}
}

@book{golub2013matrix,
  title={Matrix Computations},
  author={Golub, Gene H and Van Loan, Charles F},
  year={2013},
  edition={4th},
  publisher={Johns Hopkins University Press},
  address={Baltimore}
}

@book{trefethen1997numerical,
  title={Numerical linear algebra},
  author={Trefethen, Lloyd N and Bau, David},
  year={2022},
  publisher={SIAM}
}

@book{higham2002accuracy,
  title={Accuracy and Stability of Numerical Algorithms},
  author={Higham, Nicholas J},
  year={2002},
  edition={2nd},
  publisher={SIAM},
  address={Philadelphia}
}

@article{makridakis1979accuracy,
  title={Accuracy of forecasting: An empirical investigation},
  author={Makridakis, Spyros and Hibon, Michele},
  journal={Journal of the Royal Statistical Society: Series A (General)},
  volume={142},
  number={2},
  pages={97--125},
  year={1979},
  publisher={Wiley Online Library}
}

@book{zellner2001keep,
  title={Simplicity, inference and modelling: Keeping it sophisticatedly simple},
  author={Zellner, Arnold and Keuzenkamp, Hugo A and McAleer, Michael},
  year={2002},
  publisher={Cambridge University Press}
}

@article{green2015simple,
  title={Simple versus complex forecasting: The evidence},
  author={Green, Kesten C and Armstrong, J Scott},
  journal={Journal of Business Research},
  volume={68},
  number={8},
  pages={1678--1685},
  year={2015},
  publisher={Elsevier}
}

@article{ledoit2004well,
  title={A well-conditioned estimator for large-dimensional covariance matrices},
  author={Ledoit, Olivier and Wolf, Michael},
  journal={Journal of Multivariate Analysis},
  volume={88},
  number={2},
  pages={365--411},
  year={2004},
  publisher={Elsevier}
}

@article{bickel2008regularized,
  title={Regularized estimation of large covariance matrices},
  author={Bickel, Peter J and Levina, Elizaveta},
  journal={The Annals of Statistics},
  volume={36},
  number={1},
  pages={199--227},
  year={2008},
  publisher={Institute of Mathematical Statistics}
}

@article{ledoit2017nonlinear,
  title={Nonlinear shrinkage of the covariance matrix for portfolio selection: Markowitz meets Goldilocks},
  author={Ledoit, Olivier and Wolf, Michael},
  journal={The Review of Financial Studies},
  volume={30},
  number={12},
  pages={4349--4388},
  year={2017},
  publisher={Oxford University Press}
}

@article{denard2021factor,
  title={Factor models for portfolio selection in large dimensions: The good, the better and the ugly},
  author={De Nard, Gianluca and Ledoit, Olivier and Wolf, Michael},
  journal={Journal of Financial Econometrics},
  volume={19},
  number={2},
  pages={236--257},
  year={2021},
  publisher={Oxford University Press}
}

@article{ang2006cross,
  title={The cross-section of volatility and expected returns},
  author={Ang, Andrew and Hodrick, Robert J and Xing, Yuhang and Zhang, Xiaoyan},
  journal={The Journal of Finance},
  volume={61},
  number={1},
  pages={259--299},
  year={2006},
  publisher={Wiley Online Library}
}

@article{asness2020betting,
  title={Betting against correlation: Testing theories of the low-risk effect},
  author={Asness, Cliff and Frazzini, Andrea and Gormsen, Niels Joachim and Pedersen, Lasse Heje},
  journal={Journal of Financial Economics},
  volume={135},
  number={3},
  pages={629--652},
  year={2020},
  publisher={Elsevier}
}

@article{baker2011benchmarks,
  title={Benchmarks as limits to arbitrage: Explaining the low-volatility anomaly},
  author={Baker, Malcolm and Bradley, Brendan and Wurgler, Jeffrey},
  journal={Financial Analysts Journal},
  volume={67},
  number={1},
  pages={40--54},
  year={2011},
  publisher={Taylor \& Francis}
}

@article{magdonismail2004,
  title={Maximum drawdown},
  author={Magdon-Ismail, Malik and Atiya, Amir F},
  journal={Risk Magazine},
  volume={17},
  number={10},
  pages={99--102},
  year={2004}
}

@article{grossman1993optimal,
  title={Optimal investment strategies for controlling drawdowns},
  author={Grossman, Sanford J and Zhou, Zhongquan},
  journal={Mathematical Finance},
  volume={3},
  number={3},
  pages={241--276},
  year={1993},
  publisher={Wiley Online Library}
}

@article{kelly2014tail,
  title={Tail risk and asset prices},
  author={Kelly, Bryan and Jiang, Hao},
  journal={The Review of Financial Studies},
  volume={27},
  number={10},
  pages={2841--2871},
  year={2014},
  publisher={Oxford University Press}
}

@article{barro2021rare,
  title={Rare events and long-run risks},
  author={Barro, Robert J and Jin, Tao},
  journal={Review of economic dynamics},
  volume={39},
  pages={1--25},
  year={2021},
  publisher={Elsevier}
}

@article{michaud1989markowitz,
  title={The Markowitz optimization enigma: Is ‘optimized’optimal?},
  author={Michaud, Richard O},
  journal={Financial analysts journal},
  volume={45},
  number={1},
  pages={31--42},
  year={1989},
  publisher={Taylor \& Francis}
}

@article{jagannathan2003risk,
  title={Risk reduction in large portfolios: Why imposing the wrong constraints helps},
  author={Jagannathan, Ravi and Ma, Tongshu},
  journal={The journal of finance},
  volume={58},
  number={4},
  pages={1651--1683},
  year={2003},
  publisher={Wiley Online Library}
}

@article{chopra1993effect,
  title={The effect of errors in means, variances, and covariances on optimal portfolio choice},
  author={Chopra, Vijay K and Ziemba, William T and others},
  journal={Journal of Portfolio Management},
  volume={19},
  number={2},
  pages={6--11},
  year={1993},
  publisher={World Scientific}
}

@article{kan2007optimal,
  title={Optimal portfolio choice with parameter uncertainty},
  author={Kan, Raymond and Zhou, Guofu},
  journal={Journal of Financial and Quantitative Analysis},
  volume={42},
  number={3},
  pages={621--656},
  year={2007},
  publisher={Cambridge University Press}
}

@article{yuan2024naive,
  title={Why naive diversification is not so naive, and how to beat it?},
  author={Yuan, Ming and Zhou, Guofu},
  journal={Journal of Financial and Quantitative Analysis},
  volume={59},
  number={8},
  pages={3601--3632},
  year={2024},
  publisher={Cambridge University Press}
}

@article{xidonas2020robust,
  title={Robust portfolio optimization: a categorized bibliographic review},
  author={Xidonas, Panos and Steuer, Ralph and Hassapis, Christis},
  journal={Annals of Operations Research},
  volume={292},
  number={1},
  pages={533--552},
  year={2020},
  publisher={Springer}
}

@article{maillard2010properties,
  title={The properties of equally weighted risk contribution portfolios},
  author={Maillard, S{\'e}bastien and Roncalli, Thierry and Te{\"\i}letche, J{\'e}r{\^o}me},
  journal={Journal of portfolio management},
  volume={36},
  number={4},
  pages={60},
  year={2010}
}

@article{ararat2024mad,
  title={MAD risk parity portfolios},
  author={Ararat, {\c{C}}a{\u{g}}{\i}n and Cesarone, Francesco and P{\i}nar, Mustafa {\c{C}}elebi and Ricci, Jacopo Maria},
  journal={Annals of Operations Research},
  volume={336},
  number={1},
  pages={899--924},
  year={2024},
  publisher={Springer}
}

@article{mausser2018long,
  title={Long-only equal risk contribution portfolios for CVaR under discrete distributions},
  author={Mausser, Helmut and Romanko, Oleksandr},
  journal={Quantitative Finance},
  volume={18},
  number={11},
  pages={1927--1945},
  year={2018},
  publisher={Taylor \& Francis}
}

@book{roncalli2013introduction,
  title={Introduction to risk parity and budgeting},
  author={Roncalli, Thierry},
  year={2013},
  publisher={CRC Press}
}

@article{bruder2012managing,
  title={Managing risk exposures using the risk budgeting approach},
  author={Bruder, Benjamin and Roncalli, Thierry},
  year={2012}
}

@article{choueifaty2008toward,
  title={Toward maximum diversification},
  author={Choueifaty, Yves and Coignard, Yves},
  journal={The Journal of Portfolio Management},
  volume={35},
  number={1},
  pages={40--51},
  year={2008},
  publisher={Institutional Investor Journals Umbrella}
}

@article{yu2014diversified,
  title={Diversified portfolios with different entropy measures},
  author={Yu, Jing-Rung and Lee, Wen-Yi and Chiou, Wan-Jiun Paul},
  journal={Applied Mathematics and Computation},
  volume={241},
  pages={47--63},
  year={2014},
  publisher={Elsevier}
}

@article{ledoit2003improved,
  title={Improved estimation of the covariance matrix of stock returns with an application to portfolio selection},
  author={Ledoit, Olivier and Wolf, Michael},
  journal={Journal of empirical finance},
  volume={10},
  number={5},
  pages={603--621},
  year={2003},
  publisher={Elsevier}
}

@article{ledoit2022power,
  title={The power of (non-) linear shrinking: A review and guide to covariance matrix estimation},
  author={Ledoit, Olivier and Wolf, Michael},
  journal={Journal of Financial Econometrics},
  volume={20},
  number={1},
  pages={187--218},
  year={2022},
  publisher={Oxford University Press}
}

@book{knight1921risk,
  title={Risk, Uncertainty and Profit},
  author={Knight, Frank H.},
  year={1921},
  publisher={Houghton Mifflin},
  address={Boston, MA}
}

@article{hwang2018naive,
  title={Naive versus optimal diversification: Tail risk and performance},
  author={Hwang, Inchang and Xu, Simon and In, Francis},
  journal={European Journal of Operational Research},
  volume={265},
  number={1},
  pages={372--388},
  year={2018},
  publisher={Elsevier}
}

@book{lhabitant2017portfolio,
  title={Portfolio diversification},
  author={Lhabitant, Fran{\c{c}}ois-Serge},
  year={2017},
  publisher={Elsevier}
}

@article{koumou2020diversification,
  title={Diversification and portfolio theory: a review},
  author={Koumou, Gilles Boevi},
  journal={Financial Markets and Portfolio Management},
  volume={34},
  number={3},
  pages={267--312},
  year={2020},
  publisher={Springer Nature BV}
}

@book{mehta2004random,
  title={Random Matrices},
  author={Mehta, Madan Lal},
  volume={142},
  year={2004},
  publisher={Elsevier},
  edition={3rd},
  address={Amsterdam}
}

@article{laloux1999random,
  title={Random matrix theory and financial correlations},
  author={Laloux, Laurent and Cizeau, Pierre and Potters, Marc and Bouchaud, Jean-Philippe},
  journal={Physical Review Letters},
  volume={83},
  number={7},
  pages={1467--1470},
  year={1999},
  publisher={APS}
}

\newpage

\section*{Appendix} \label{app}

In this appendix the tables resuming the results for the Nikkei and FTSE indexes are reported.

%%%%%%%%%%%%%%%%%%%%%%%%%%%%%%%%%%%%%%%%%%%%%%%%%%%%%%%%%%%%%%%%%%%%%%%%%%%%%%%%%%%%%%%%%%%%%%%%%%%%%%%%%%%%%%%%%%%%%%%%%%%%%%%%%%%
%%%%%%%% tabelle Nikkei %%%%%%%%%%%%%%%%%%%%%%%%%%%%%%%%%%%%%%%%%%%%%%%%%%%%%%%%%%%%%%%%%%%%%%%%%%%%%%%%%%%%%%%%%%%%%%%%%%%%%%%%%%%

\begin{table}[!h]
  \centering
  \caption{Comparison between the out-of-sample returns of the investment strategies for different values of $N = 5, 10, 20$ and  $w = 20$. Database: Nikkei.}
	\label{tabnikkei1}%
	\begin{adjustbox}{width=\textwidth}
    \begin{tabular}{|c|l|rrrrrr|}
    \toprule
    \multicolumn{1}{|r}{} &       & \multicolumn{6}{c|}{w = 20} \\
    \multicolumn{1}{|r}{} &       & \multicolumn{1}{c}{\textbf{1/N}} & \multicolumn{1}{c}{\textbf{RR}} & \multicolumn{1}{c}{\textbf{random}} & \multicolumn{1}{c}{\textbf{Min-var}} & \multicolumn{1}{c}{\textbf{Min-VaR}} & \multicolumn{1}{c|}{\textbf{Min-CVaR}} \\
    \midrule
    \multirow{8}[2]{*}{N = 5} & \textbf{a.r.} & 9.17E-05 & 9.87E-06 & -1.56E-04 & -0.00012 & -9.50E-05 & -9.71E-05 \\
          & \textbf{st. dev.} & 0.016512 & 0.009988 & 0.01093 & 0.015482 & 0.016081 & 0.016412 \\
          & \textbf{SR} & 0.005626 & 0.00103 & -0.01421 & -0.00768 & -0.00596 & -0.00584 \\
          & \textbf{VaR 1\%} & 0.046418 & 0.029513 & 0.032229 & 0.043337 & 0.04518 & 0.046703 \\
          & \textbf{CVaR 1\%} & 0.066752 & 0.039959 & 0.046373 & 0.063184 & 0.064231 & 0.065718 \\
          & \textbf{MDD} & 0.680658 & 0.542736 & 0.66358 & 0.729285 & 0.76124 & 0.760502 \\
          & \textbf{Sk } & -0.35233 & -0.41984 & -0.35694 & -0.37914 & -0.29529 & -0.3 \\
          & \textbf{K } & 10.30454 & 8.354047 & 15.58827 & 11.36062 & 10.05021 & 10.62444 \\
    \midrule
    \multirow{8}[2]{*}{N = 10} & \textbf{a.r.} & 1.26E-05 & -3.65E-05 & -1.93E-04 & -1.68E-04 & -1.92E-04 & -0.00017 \\
          & \textbf{st. dev.} & 0.016327 & 0.009665 & 0.010786 & 0.0141 & 0.014938 & 0.015208 \\
          & \textbf{SR} & 0.000841 & -0.00379 & -0.01794 & -0.01181 & -0.01274 & -0.01139 \\
          & \textbf{VaR 1\%} & 0.045837 & 0.028937 & 0.03258 & 0.039524 & 0.042893 & 0.043682 \\
          & \textbf{CVaR 1\%} & 0.066662 & 0.039048 & 0.047327 & 0.057426 & 0.060714 & 0.06199 \\
          & \textbf{MDD} & 0.686758 & 0.553315 & 0.69014 & 0.757775 & 0.773431 & 0.776967 \\
          & \textbf{Sk } & -0.35806 & -0.45454 & -0.69785 & -0.44875 & -0.3977 & -0.38568 \\
          & \textbf{K } & 9.872222 & 7.924353 & 14.3251 & 11.84791 & 11.62733 & 11.17076 \\
    \midrule
    \multirow{8}[2]{*}{N = 20} & \textbf{a.r.} & 2.52E-05 & -4.21E-05 & -1.72E-04 & -2.43E-04 & -2.30E-04 & -0.00016 \\
          & \textbf{st. dev.} & 0.015774 & 0.009494 & 0.010807 & 0.013094 & 0.014041 & 0.01427 \\
          & \textbf{SR} & 0.001663 & -0.00453 & -0.01594 & -0.01848 & -0.01645 & -0.01127 \\
          & \textbf{VaR 1\%} & 0.044455 & 0.028536 & 0.032763 & 0.036951 & 0.040304 & 0.041442 \\
          & \textbf{CVaR 1\%} & 0.064839 & 0.039394 & 0.047506 & 0.053807 & 0.057084 & 0.057674 \\
          & \textbf{MDD} & 0.671452 & 0.552684 & 0.645322 & 0.791478 & 0.807554 & 0.772175 \\
          & \textbf{Sk } & -0.37694 & -0.57273 & -0.52298 & -0.51186 & -0.48009 & -0.45163 \\
          & \textbf{K } & 10.31537 & 8.510419 & 16.5101 & 11.23438 & 9.856402 & 9.543052 \\
    \bottomrule
    \end{tabular}%
  \end{adjustbox}
\end{table}%

\begin{table}[!h]
  \centering
  \caption{Comparison between the out-of-sample returns of the investment strategies for different values of $N = 5, 10, 20$ and  $w = 30$. Database: Nikkei.}
	\label{tabnikkei2}%
	\begin{adjustbox}{width=\textwidth}
    \begin{tabular}{|c|l|rrrrrr|}
    \toprule
    \multicolumn{1}{|r}{} &       & \multicolumn{6}{c|}{w = 30} \\
    \multicolumn{1}{|r}{} &       & \multicolumn{1}{c}{\textbf{1/N}} & \multicolumn{1}{c}{\textbf{RR}} & \multicolumn{1}{c}{\textbf{random}} & \multicolumn{1}{c}{\textbf{Min-var}} & \multicolumn{1}{c}{\textbf{Min-VaR}} & \multicolumn{1}{c|}{\textbf{Min-CVaR}} \\
    \midrule
    \multirow{8}[2]{*}{N = 5} & \textbf{a.r.} & 5.81E-05 & -2.41E-05 & -1.55E-04 & -4.51E-05 & -3.56E-05 & -3.72E-05 \\
          & \textbf{st. dev.} & 0.017263 & 0.010749 & 0.011925 & 0.016266 & 0.017035 & 0.017042 \\
          & \textbf{SR} & 0.003514 & -0.00199 & -0.01277 & -0.00226 & -0.002 & -0.00196 \\
          & \textbf{VaR 1\%} & 0.048041 & 0.031231 & 0.035139 & 0.046316 & 0.047116 & 0.047657 \\
          & \textbf{CVaR 1\%} & 0.068855 & 0.042232 & 0.050142 & 0.065386 & 0.066598 & 0.066805 \\
          & \textbf{MDD} & 0.721917 & 0.592573 & 0.692712 & 0.706337 & 0.73205 & 0.726371 \\
          & \textbf{Sk } & -0.34741 & -0.41302 & -0.44525 & -0.30279 & -0.26546 & -0.21997 \\
          & \textbf{K } & 9.944842 & 8.311775 & 14.0865 & 10.60612 & 9.68799 & 9.951139 \\
    \midrule
    \multirow{8}[2]{*}{N = 10} & \textbf{a.r.} & 8.99E-05 & 1.22E-05 & -1.74E-04 & 3.63E-06 & -7.37E-05 & -1.79E-05 \\
          & \textbf{st. dev.} & 0.016296 & 0.010167 & 0.011396 & 0.013884 & 0.015153 & 0.014965 \\
          & \textbf{SR} & 0.005533 & 0.001194 & -0.01535 & 0.000448 & -0.00449 & -0.00092 \\
          & \textbf{VaR 1\%} & 0.045604 & 0.030018 & 0.034025 & 0.038757 & 0.042798 & 0.042118 \\
          & \textbf{CVaR 1\%} & 0.066185 & 0.040538 & 0.048846 & 0.056379 & 0.062279 & 0.061112 \\
          & \textbf{MDD} & 0.67903 & 0.572938 & 0.683696 & 0.680495 & 0.735552 & 0.71234 \\
          & \textbf{Sk } & -0.39672 & -0.48995 & -0.4621 & -0.38232 & -0.471 & -0.44248 \\
          & \textbf{K } & 9.961985 & 7.800605 & 14.46874 & 12.86445 & 11.70589 & 13.12057 \\
    \midrule
    \multirow{8}[2]{*}{N = 20} & \textbf{a.r.} & 7.16E-05 & -2.19E-05 & -1.20E-04 & -4.65E-05 & -1.37E-04 & -9.36E-05 \\
          & \textbf{st. dev.} & 0.015826 & 0.009135 & 0.010227 & 0.012887 & 0.014493 & 0.013952 \\
          & \textbf{SR} & 0.004606 & -0.0022 & -0.01145 & -0.00354 & -0.00932 & -0.00669 \\
          & \textbf{VaR 1\%} & 0.04373 & 0.027949 & 0.031225 & 0.036116 & 0.041215 & 0.040361 \\
          & \textbf{CVaR 1\%} & 0.06543 & 0.038714 & 0.04393 & 0.052324 & 0.060231 & 0.057172 \\
          & \textbf{MDD} & 0.666782 & 0.527051 & 0.58294 & 0.634816 & 0.747071 & 0.714909 \\
          & \textbf{Sk } & -0.39706 & -0.60712 & -0.30645 & -0.42976 & -0.53715 & -0.47204 \\
          & \textbf{K } & 10.48243 & 9.308393 & 16.43231 & 11.62153 & 13.48451 & 10.9378 \\
    \bottomrule
    \end{tabular}%
  \end{adjustbox}
\end{table}%

\begin{table}[!h]
  \centering
  \caption{Comparison between the out-of-sample returns of the investment strategies for different values of $N = 5, 10, 20$ and  $w = 40$. Database: Nikkei.}
	\label{tabnikkei3}%
	\begin{adjustbox}{width=\textwidth}
    \begin{tabular}{|c|l|rrrrrr|}
    \toprule
    \multicolumn{1}{|r}{} &       & \multicolumn{6}{c|}{w = 40} \\
    \multicolumn{1}{|r}{} &       & \multicolumn{1}{c}{\textbf{1/N}} & \multicolumn{1}{c}{\textbf{RR}} & \multicolumn{1}{c}{\textbf{random}} & \multicolumn{1}{c}{\textbf{Min-var}} & \multicolumn{1}{c}{\textbf{Min-VaR}} & \multicolumn{1}{c|}{\textbf{Min-CVaR}} \\
    \midrule
    \multirow{8}[2]{*}{N = 5} & \textbf{a.r.} & 6.28E-05 & -1.11E-05 & -1.27E-04 & -2.89E-05 & -9.61E-05 & -6.08E-05 \\
          & \textbf{st. dev.} & 0.017137 & 0.010111 & 0.011019 & 0.015642 & 0.016552 & 0.016625 \\
          & \textbf{SR} & 0.003696 & -0.00098 & -0.01131 & -0.0018 & -0.00577 & -0.00363 \\
          & \textbf{VaR 1\%} & 0.04742 & 0.02977 & 0.032525 & 0.044597 & 0.046753 & 0.047367 \\
          & \textbf{CVaR 1\%} & 0.069381 & 0.040754 & 0.046546 & 0.064951 & 0.06835 & 0.068208 \\
          & \textbf{MDD} & 0.691864 & 0.542401 & 0.598903 & 0.685528 & 0.728924 & 0.717413 \\
          & \textbf{Sk } & -0.39539 & -0.45565 & -0.40832 & -0.5415 & -0.4323 & -0.39098 \\
          & \textbf{K } & 9.544987 & 8.326782 & 12.83043 & 11.76469 & 11.25995 & 11.25003 \\
    \midrule
    \multirow{8}[2]{*}{N = 10} & \textbf{a.r.} & 0.000101 & 3.23E-05 & -1.59E-04 & -1.81E-05 & -5.00E-05 & -3.55E-05 \\
          & \textbf{st. dev.} & 0.015812 & 0.009558 & 0.010652 & 0.01351 & 0.014461 & 0.014492 \\
          & \textbf{SR} & 0.00636 & 0.003412 & -0.01495 & -0.00125 & -0.00348 & -0.00244 \\
          & \textbf{VaR 1\%} & 0.044211 & 0.028194 & 0.032472 & 0.037651 & 0.040508 & 0.040699 \\
          & \textbf{CVaR 1\%} & 0.063936 & 0.038314 & 0.045468 & 0.055378 & 0.058699 & 0.059131 \\
          & \textbf{MDD} & 0.662363 & 0.515413 & 0.645074 & 0.670019 & 0.707117 & 0.691282 \\
          & \textbf{Sk } & -0.34898 & -0.42233 & -0.40773 & -0.42288 & -0.43763 & -0.44026 \\
          & \textbf{K } & 10.49599 & 8.332075 & 13.60905 & 12.75198 & 11.87922 & 11.80406 \\
    \midrule
    \multirow{8}[2]{*}{N = 20} & \textbf{a.r.} & 7.22E-05 & 1.97E-05 & -1.24E-04 & -6.91E-05 & -0.00011 & -8.61E-05 \\
          & \textbf{st. dev.} & 0.015751 & 0.008848 & 0.009847 & 0.012681 & 0.014004 & 0.013879 \\
          & \textbf{SR} & 0.004622 & 0.002298 & -0.01252 & -0.00528 & -0.00744 & -0.00607 \\
          & \textbf{VaR 1\%} & 0.044545 & 0.026136 & 0.029488 & 0.036198 & 0.040337 & 0.040131 \\
          & \textbf{CVaR 1\%} & 0.064928 & 0.0367 & 0.042653 & 0.053243 & 0.058939 & 0.058114 \\
          & \textbf{MDD} & 0.66619 & 0.504408 & 0.567433 & 0.645689 & 0.721354 & 0.705878 \\
          & \textbf{Sk } & -0.37342 & -0.52896 & -0.41222 & -0.48903 & -0.52211 & -0.49469 \\
          & \textbf{K } & 10.97731 & 9.740387 & 16.67813 & 12.14177 & 12.03183 & 11.39723 \\
    \bottomrule
    \end{tabular}%
  \end{adjustbox}%
\end{table}%

%%%%%%%%%%%%%%%%%%%%%%%%%%%%%%%%%%%%%%%%%%%%%%%%%%%%%%%%%%%%%%%%%%%%%%%%%%%%%%%%%%%%%%%%%%%%%%%%%%%%%%%%%%%%%%%%%%%%%%%%%%%%%%%%%%%
%%%%%% tabelle ftse%%%%%%%%%%%%%%%%%%%%%%%%%%%%%%%%%%%%%%%%%%%%%%%%%%%%%%%%%%%%%%%%%%%%%%%%%%%%%%%%%%%%%%%%%%%%%%%%%%%%%%%%%%%%%%%%

\begin{table}[!h]
  \centering
  \caption{Comparison between the out-of-sample returns of the investment strategies for different values of $N = 5, 10, 20$ and  $w = 20$. Database: FTSE.}
	\label{tabftse1}%
	\begin{adjustbox}{width=\textwidth}
    \begin{tabular}{|c|l|rrrrrr|}
    \toprule
    \multicolumn{1}{|r}{} &       & \multicolumn{6}{c|}{w = 20} \\
    \multicolumn{1}{|r}{} &       & \multicolumn{1}{c}{\textbf{1/N}} & \multicolumn{1}{c}{\textbf{RR}} & \multicolumn{1}{c}{\textbf{random}} & \multicolumn{1}{c}{\textbf{Min-var}} & \multicolumn{1}{c}{\textbf{Min-VaR}} & \multicolumn{1}{c|}{\textbf{Min-CVaR}} \\
    \midrule
    \multirow{8}[2]{*}{N = 5} & \textbf{a.r.} & 2.06E-04 & 1.10E-04 & -9.15E-05 & 5.30E-05 & 1.49E-05 & 4.11E-05 \\
          & \textbf{st. dev.} & 0.014366 & 0.008961 & 0.010202 & 0.013436 & 0.013961 & 0.014186 \\
          & \textbf{SR} & 0.014322 & 0.012202 & -0.00914 & 0.00434 & 0.001572 & 0.003391 \\
          & \textbf{VaR 1\%} & 0.04305 & 0.026334 & 0.031189 & 0.039609 & 0.040754 & 0.041507 \\
          & \textbf{CVaR 1\%} & 0.060277 & 0.035277 & 0.044783 & 0.055886 & 0.057506 & 0.058796 \\
          & \textbf{MDD} & 0.642536 & 0.478197 & 0.577239 & 0.645721 & 0.660804 & 0.645366 \\
          & \textbf{Sk } & -0.4828 & -0.31091 & -0.56961 & -0.71108 & -0.5763 & -0.50989 \\
          & \textbf{K } & 10.30071 & 8.053948 & 14.79245 & 15.51227 & 15.40541 & 14.23376 \\
    \midrule
    \multirow{8}[2]{*}{N = 10} & \textbf{a.r.} & 2.03E-04 & 9.01E-05 & -1.18E-04 & 1.18E-05 & 2.88E-06 & 2.40E-05 \\
          & \textbf{st. dev.} & 0.013162 & 0.00864 & 0.010169 & 0.011278 & 0.011794 & 0.011927 \\
          & \textbf{SR} & 0.015571 & 0.010278 & -0.01145 & 0.001092 & 0.000269 & 0.002069 \\
          & \textbf{VaR 1\%} & 0.038308 & 0.025134 & 0.031587 & 0.031732 & 0.033554 & 0.034439 \\
          & \textbf{CVaR 1\%} & 0.053799 & 0.032872 & 0.044291 & 0.04563 & 0.046394 & 0.047728 \\
          & \textbf{MDD} & 0.591282 & 0.459185 & 0.580667 & 0.550456 & 0.573337 & 0.59257 \\
          & \textbf{Sk } & -0.42719 & -0.2715 & -0.49752 & -0.56494 & -0.32385 & -0.42935 \\
          & \textbf{K } & 10.17227 & 7.943349 & 14.99898 & 11.82135 & 10.09322 & 10.66026 \\
    \midrule
    \multirow{8}[2]{*}{N = 20} & \textbf{a.r.} & 1.63E-04 & 6.35E-05 & -9.34E-05 & -1.03E-04 & -9.86E-05 & -3.96E-05 \\
          & \textbf{st. dev.} & 0.01229 & 0.007843 & 0.009533 & 0.010465 & 0.010815 & 0.010957 \\
          & \textbf{SR} & 0.013466 & 0.008099 & -0.00997 & -0.0099 & -0.00915 & -0.00359 \\
          & \textbf{VaR 1\%} & 0.036176 & 0.023072 & 0.029754 & 0.029578 & 0.031038 & 0.031911 \\
          & \textbf{CVaR 1\%} & 0.051258 & 0.029908 & 0.041295 & 0.043496 & 0.043688 & 0.045035 \\
          & \textbf{MDD} & 0.555352 & 0.40234 & 0.551438 & 0.604526 & 0.599253 & 0.56185 \\
          & \textbf{Sk } & -0.46686 & -0.39455 & -0.37477 & -0.54759 & -0.40257 & -0.50966 \\
          & \textbf{K } & 11.07022 & 7.141955 & 14.66003 & 13.69527 & 10.94692 & 11.66037 \\
    \bottomrule
    \end{tabular}%
  \end{adjustbox}%
\end{table}%

\begin{table}[!h]
  \centering
  \caption{Comparison between the out-of-sample returns of the investment strategies for different values of $N = 5, 10, 20$ and  $w = 30$. Database: FTSE.}
	\label{tabftse2}%
	\begin{adjustbox}{width=\textwidth}
    \begin{tabular}{|c|l|rrrrrr|}
    \toprule
    \multicolumn{1}{|r}{} &       & \multicolumn{6}{c|}{w = 30} \\
    \multicolumn{1}{|r}{} &       & \multicolumn{1}{c}{\textbf{1/N}} & \multicolumn{1}{c}{\textbf{RR}} & \multicolumn{1}{c}{\textbf{random}} & \multicolumn{1}{c}{\textbf{Min-var}} & \multicolumn{1}{c}{\textbf{Min-VaR}} & \multicolumn{1}{c|}{\textbf{Min-CVaR}} \\
    \midrule
    \multirow{8}[2]{*}{N = 5} & \textbf{a.r.} & 1.40E-04 & 5.70E-05 & -1.57E-04 & 9.38E-05 & 2.20E-05 & 7.14E-05 \\
          & \textbf{st. dev.} & 0.013645 & 0.008729 & 0.01019 & 0.012084 & 0.01291 & 0.012712 \\
          & \textbf{SR} & 0.010455 & 0.00671 & -0.01558 & 0.008259 & 0.002123 & 0.005999 \\
          & \textbf{VaR 1\%} & 0.039798 & 0.025208 & 0.031478 & 0.033742 & 0.035992 & 0.03522 \\
          & \textbf{CVaR 1\%} & 0.057365 & 0.034972 & 0.045621 & 0.050085 & 0.053219 & 0.052001 \\
          & \textbf{MDD} & 0.615466 & 0.453587 & 0.632739 & 0.517488 & 0.587764 & 0.544679 \\
          & \textbf{Sk } & -0.46524 & -0.349 & -0.57323 & -0.65181 & -0.68574 & -0.77134 \\
          & \textbf{K } & 11.05291 & 8.879098 & 16.16173 & 14.39763 & 24.291 & 21.34112 \\
    \midrule
    \multirow{8}[2]{*}{N = 10} & \textbf{a.r.} & 1.65E-04 & 6.83E-05 & -1.13E-04 & 2.83E-05 & 4.08E-05 & 5.32E-05 \\
          & \textbf{st. dev.} & 0.012705 & 0.008426 & 0.009986 & 0.01094 & 0.011455 & 0.011362 \\
          & \textbf{SR} & 0.013007 & 0.007875 & -0.01167 & 0.002455 & 0.003459 & 0.0047 \\
          & \textbf{VaR 1\%} & 0.037344 & 0.024527 & 0.030193 & 0.030355 & 0.032544 & 0.031858 \\
          & \textbf{CVaR 1\%} & 0.052226 & 0.031643 & 0.043636 & 0.044462 & 0.045532 & 0.045121 \\
          & \textbf{MDD} & 0.574816 & 0.429127 & 0.58659 & 0.542314 & 0.579408 & 0.538715 \\
          & \textbf{Sk } & -0.46118 & -0.34995 & -0.40035 & -0.42863 & -0.4404 & -0.30208 \\
          & \textbf{K } & 10.5078 & 6.821582 & 13.25751 & 12.74558 & 10.42009 & 11.37505 \\
    \midrule
    \multirow{8}[2]{*}{N = 20} & \textbf{a.r.} & 1.67E-04 & 3.43E-05 & -1.28E-04 & -5.66E-05 & -1.64E-05 & 1.22E-06 \\
          & \textbf{st. dev.} & 0.012515 & 0.007824 & 0.009621 & 0.010303 & 0.011008 & 0.010946 \\
          & \textbf{SR} & 0.013317 & 0.004434 & -0.01338 & -0.00555 & -0.00158 & 8.87E-05 \\
          & \textbf{VaR 1\%} & 0.037238 & 0.023253 & 0.03034 & 0.030132 & 0.032307 & 0.032162 \\
          & \textbf{CVaR 1\%} & 0.052787 & 0.030368 & 0.043705 & 0.04276 & 0.045575 & 0.044726 \\
          & \textbf{MDD} & 0.569901 & 0.412761 & 0.570488 & 0.57126 & 0.584957 & 0.568598 \\
          & \textbf{Sk } & -0.4812 & -0.44393 & -0.61361 & -0.56517 & -0.62106 & -0.41623 \\
          & \textbf{K } & 10.99117 & 7.267333 & 17.01287 & 12.70416 & 11.61648 & 11.10315 \\
    \bottomrule
    \end{tabular}%
  \end{adjustbox}%
\end{table}%

\begin{table}[!h]
  \centering
  \caption{Comparison between the out-of-sample returns of the investment strategies for different values of $N = 5, 10, 20$ and  $w = 40$. Database: FTSE.}
	\label{tabftse3}%
	\begin{adjustbox}{width=\textwidth}
    \begin{tabular}{|c|l|rrrrrr|}
    \toprule
    \multicolumn{1}{|r}{} &       & \multicolumn{6}{c|}{w = 40} \\
    \multicolumn{1}{|r}{} &       & \multicolumn{1}{c}{\textbf{1/N}} & \multicolumn{1}{c}{\textbf{RR}} & \multicolumn{1}{c}{\textbf{random}} & \multicolumn{1}{c}{\textbf{Min-var}} & \multicolumn{1}{c}{\textbf{Min-VaR}} & \multicolumn{1}{c|}{\textbf{Min-CVaR}} \\
    \midrule
    \multirow{8}[2]{*}{N = 5} & \textbf{a.r.} & 2.10E-04 & 1.26E-04 & -1.30E-04 & 1.43E-04 & 8.29E-05 & 1.25E-04 \\
          & \textbf{st. dev.} & 0.013022 & 0.008793 & 0.010114 & 0.011891 & 0.012376 & 0.012488 \\
          & \textbf{SR} & 0.015958 & 0.013856 & -0.01296 & 0.011878 & 0.006657 & 0.009858 \\
          & \textbf{VaR 1\%} & 0.037814 & 0.024797 & 0.030539 & 0.034311 & 0.035285 & 0.035691 \\
          & \textbf{CVaR 1\%} & 0.052708 & 0.033186 & 0.043421 & 0.0476 & 0.04979 & 0.049991 \\
          & \textbf{MDD} & 0.532464 & 0.41596 & 0.587355 & 0.527314 & 0.554256 & 0.552075 \\
          & \textbf{Sk } & -0.46744 & -0.31439 & -0.61898 & -0.47873 & -0.51812 & -0.45788 \\
          & \textbf{K } & 9.528321 & 7.45499 & 13.59549 & 11.18646 & 12.16766 & 11.21791 \\
    \midrule
    \multirow{8}[2]{*}{N = 10} & \textbf{a.r.} & 0.000198 & 1.12E-04 & -8.06E-05 & 7.21E-05 & 4.35E-05 & 5.87E-05 \\
          & \textbf{st. dev.} & 0.013001 & 0.008395 & 0.010188 & 0.010691 & 0.011483 & 0.011446 \\
          & \textbf{SR} & 0.015392 & 0.013385 & -0.00794 & 0.006802 & 0.003822 & 0.004999 \\
          & \textbf{VaR 1\%} & 0.0379 & 0.024009 & 0.030156 & 0.029855 & 0.031823 & 0.031958 \\
          & \textbf{CVaR 1\%} & 0.053405 & 0.031723 & 0.044204 & 0.043133 & 0.046757 & 0.046164 \\
          & \textbf{MDD} & 0.557649 & 0.419382 & 0.546096 & 0.483133 & 0.540006 & 0.520942 \\
          & \textbf{Sk } & -0.41787 & -0.35819 & -0.46959 & -0.38761 & -0.49886 & -0.43899 \\
          & \textbf{K } & 10.54285 & 6.92313 & 15.88459 & 11.6566 & 14.33842 & 13.33959 \\
    \midrule
    \multirow{8}[2]{*}{N = 20} & \textbf{a.r.} & 1.51E-04 & 5.84E-05 & -1.47E-04 & 2.66E-05 & 1.93E-05 & 2.44E-05 \\
          & \textbf{st. dev.} & 0.012125 & 0.007674 & 0.009278 & 0.010003 & 0.010609 & 0.010546 \\
          & \textbf{SR} & 0.012564 & 0.007652 & -0.01587 & 0.002667 & 0.00179 & 0.002476 \\
          & \textbf{VaR 1\%} & 0.036251 & 0.022866 & 0.02862 & 0.028647 & 0.03016 & 0.030446 \\
          & \textbf{CVaR 1\%} & 0.050553 & 0.029478 & 0.041538 & 0.041349 & 0.04269 & 0.042901 \\
          & \textbf{MDD} & 0.579307 & 0.416526 & 0.595656 & 0.487407 & 0.510753 & 0.500245 \\
          & \textbf{Sk } & -0.47824 & -0.43374 & -0.6544 & -0.45144 & -0.41739 & -0.52681 \\
          & \textbf{K } & 10.48219 & 6.855555 & 15.98496 & 12.27365 & 11.59038 & 10.47927 \\
    \bottomrule
    \end{tabular}%
  \end{adjustbox}%
\end{table}%

\end{document}